\documentclass{llncs}
\usepackage{makeidx}  
\usepackage{mathtools} 
\usepackage{pgfplots} 
\usepackage{bibgerm} 
\usepackage[english]{babel} 
\usepackage{textcomp} 
\usepackage{hyperref}
\usepackage{authblk}

\definecolor{tudolight}{rgb}{0.91,0.97,0.76}

\newcommand{\qgrams}{\mathcal{Q}}

\pgfplotsset{mystyle/.style={thick, smooth, mark size = 3pt}}
\pgfplotsset{barstyle/.style={black}}

\pgfplotsset{styleBWA/.style={mystyle, color=red, mark=x}}
\pgfplotsset{styleBWAedit/.style={mystyle, color=orange, mark=+}}
\pgfplotsset{styleBWAsw/.style={mystyle, color=orange!60!black, mark=asterisk}}
\pgfplotsset{styleBowtie/.style={mystyle, color=blue, mark=star}}
\pgfplotsset{styleVATRAM/.style={mystyle, color=green, mark=Mercedes star}}
\pgfplotsset{styleVATRAMnovar/.style={mystyle, color=green!75!cyan!50!black, mark=Mercedes star flipped}}
\pgfplotsset{styleMRSfast/.style={mystyle, color=cyan, mark=|}}

\pgfplotsset{barstyleBWA/.style={barstyle, fill=red}}
\pgfplotsset{barstyleBWAedit/.style={barstyle, fill=orange}}
\pgfplotsset{barstyleBWAsw/.style={barstyle, fill=orange!60!black}}
\pgfplotsset{barstyleBowtie/.style={barstyle, fill=blue}}
\pgfplotsset{barstyleVATRAM/.style={barstyle, fill=green}}
\pgfplotsset{barstyleVATRAMnovar/.style={barstyle, fill=green!75!cyan!50!black}}
\pgfplotsset{barstyleMRSfast/.style={barstyle, fill=cyan}}
\pgfplotsset{barstyleRazer/.style={barstyle, fill=black!35!white}}
\pgfplotsset{barstyleRazerRabema/.style={barstyle, fill=black!65!white}}

\begin{document}
\pagestyle{headings}  

\title{Variant tolerant read mapping using min-hashing}
\author{Jens Quedenfeld\inst{1,3} \and Sven Rahmann\inst{2,3}}
\authorrunning{Jens Quedenfeld \and Sven Rahmann}
\tocauthor{Jens Quedenfeld and Sven Rahmann}
%
\institute{Chair of Theoretical Computer Science, Technical University of Munich, Germany\\
\email{jens.quedenfeld@in.tum.de}
\and
Genome Informatics, Institute of Human Genetics, University Hospital Essen, University of Duisburg-Essen, Essen, Germany\\
\email{Sven.Rahmann@uni-due.de}
\and
Bioinformatics, Computer Science XI, TU Dortmund, Germany
}
\maketitle

\begin{abstract}
DNA read mapping is a ubiquitous task in bioinformatics, and many tools have been developed to solve the read mapping problem.
However, there are two trends that are changing the landscape of readmapping:
First, new sequencing technologies provide very long reads with high error rates (up to 15\%).
Second, many genetic variants in the population are known, so the reference genome is not considered as a single string over ACGT, but as a complex object containing these variants. Most existing read mappers do not handle these new circumstances appropriately.

We introduce a new read mapper prototype called VATRAM that considers variants.
It is based on \emph{Min-Hashing} of $q$-gram sets of reference genome windows.
Min-Hashing is one form of \emph{locality sensitive hashing}.
The variants are directly inserted into VATRAMs index which leads to a fast mapping process.
Our results show that VATRAM achieves better precision and recall than state-of-the-art read mappers like BWA under certain cirumstances.
VATRAM is open source and can be accessed at \url{https://bitbucket.org/Quedenfeld/vatram-src/}.
\end{abstract}

\section{Introduction}
\label{sec:introduction}

In bioinformatics, DNA read mapping has become an important basic step for many sequencing analysis tasks.
Given millions of short DNA fragments (so called \emph{reads}) over the alphabet $\Sigma = \{A, C, G, T\}$  and a reference genome which is a long DNA string that is many magnitudes longer than the reads, the problem is to locate these reads in that genome.
A typical (short) read length is 100--300 nucleotides \cite{Metzker2010}, the human reference genome has a length of approximately 3~billion nucleotides.
Since not all individuals are identical, there are some differences between a given read and the corresponding interval on the reference genome.
Furthermore, the sequencing machines are not perfect and produces sequencing errors, therefore, the best match is searched for, e.g. according to the edit distance.

To solve the read mapping problem efficiently, usually an index data structure is used to find an exact match between a short substring of the read and the reference genome.
Afterwards, the exact match (called \emph{seed}) is extended to a full alignment.
Many popular read mappers (such as BWA \cite{BWA2009} or Bowtie2 \cite{Bowtie2012}) use the FM index which is based on suffix arrays and the Burrows Wheeler Transformation \cite{fmindex}.
However, if there are many differences between the read and the reference genome, then there are few unique long seeds that can be found, but many unspecific short seeds, so common read mappers have difficulties to map reads with many errors efficiently

This problem becomes more and more important for two reasons.
First, there are new sequencing technologies (for example from Pacific Biosciences \cite{PacBio2009}) which produce long reads with up to $60\,000$ nucleotides and a high error rate of 10\%--15\%.
Second, thousands of individual human genomes have been sequenced in the last years, so many frequent variants are known.
It is commonly accepted that the human genome is not represented well by a single string over the alphabet $\Sigma$.
To improve the read mapping process it is necessary to consider these known variants.
About 90\% of the differences between the individuals of one species are \emph{single nucleotide polymorphisms} (SNPs), which are substitutions of one single nucleotide in DNA. Other important variant types are insertions or deletions of one or more nucleotides as well as large structural rearrangements.
Index data structures based on the FM index have problems to handle these new circumstances, so new types of indexes have to be explored.

One such alternative is \emph{locality sensitive hashing} (LSH) on $q$-gram sets of reference intervals.
LSH has been already used for finding similarities between different DNA sequences \cite{Buhler2001} as well as for genome assembly \cite{Berlin2015} of PacBio reads.
However, as far as we know LSH was not used for read mapping yet.
We have developed a prototype of a read mapper called VATRAM (\emph{VAriant Tolerant ReAd Mapper}) which is able to consider known variants.
VATRAM uses \emph{Min-Hashing} \cite{Broder1997} which is one specific form of LSH.

This paper is organized as follows:
In section~\ref{sec:index} we describe in detail how the index of \textsc{VATRAM} works and how it is created.
Afterwards (section~\ref{sec:read}) we explain how reads can be found in the reference genome using this index and how they are aligned using our variant tolerant aligner.
Section~\ref{sec:exp} contains several experiments where VATRAM is compared to other read mappers.

Preliminary ideas, implementations and experiments were reported in an internal report \cite{PG583} and a Master's thesis \cite{MA}; the present article contains our summarized findings.
The source code of VATRAM is available at \url{https://bitbucket.org/Quedenfeld/vatram-src/}.

\section{Index creation}
\label{sec:index}

\subsection{Basic idea}

The index of VATRAM is created as follows.
First, each chromosome is divided into windows of length~$w$, such that the windows are slightly longer than the typical read length~$n$.
The distance between two consecutive windows is denoted by $o \leq w$, so the windows may (and usually do) overlap.
A typical configuration is $w = 1.4n$ and $o=1.25n$ \cite{MA}.

For each window, the set of contained $q$-grams (substrings of length~$q$) is determined.
If the window contains a SNP which appears with a known population frequency higher than $\delta$, then the $q$-grams containing the SNP are added to the window's $q$-gram set.
A typical value is $\delta = 0.2$.

If there are two or more SNPs with a distance lower than~$q$, then we add all combinations of $q$-grams to the set.
However, if are too many SNPs within a given substring of length $q$, than these SNPs are ignored, because adding all combinations would enlarge the $q$-gram set too much.
In the extreme case there would be $q$~consecutive positions where all four bases (A,C,G,T) are allowed.
So there would be $4^q$ possible $q$-grams for that substring.
If we added these to the window's $q$-gram set, it would contain all possible $q$-grams, so there is no information which could help us to map a read to this window.
By default a limit of $l = 3$ $q$-grams for each $q$-gram position is used.

Each $q$-gram set is mapped to a single value using a technique called min-hashing \cite{Broder1997}.
For that purpose we conceptually choose an arbitrary permutation of all $q$-grams uniformly at random.
A $q$-gram set is mapped to the smallest $q$-gram according to the order defined by this permutation.
The resulting $q$-gram is called the \emph{signature value} of the window.
If we compare two $q$-gram sets $Q$ and $Q'$ the min hash property holds:

\begin{lemma}[Min-hash property]
\label{lemma:minhash}
Let $\qgrams$ be the set of all strings over $\Sigma$ of length $q$. Given two sets $Q\subset\qgrams$, $Q'\subset\qgrams$ and the set $\Pi$ of all permutations on $\qgrams$, let $\pi \in \Pi$ be a random permutation.
For any $Q\subset\qgrams$, define $h(Q) := \min_\pi(Q)$.
The probability that $Q$ and $Q'$ are hashed to the same value $h(Q)=h(Q')$ is equal to the Jaccard coefficient of $Q$ and $Q'$,
\begin{equation}
P(h(Q)=h(Q')) = \frac{\vert Q \cap Q' \vert}{\vert Q \cup Q' \vert} .
\label{eq:minhash}
\end{equation}
\end{lemma}
A proof can be found in \cite{Broder1997}. 

Of course, in practice it is not possible to do this, because there are $(4^q)!$ different permutations, so $\log_2 (4^q!)$ bits are required to represent a permutation.
For a common value like $q=16$, this is about $4 \cdot 10^{10}$ bits per permutation, which is not practical.

Therefore, instead of a permutation we choose a random 32 bit word which is called \emph{permutation value}.
The $q$-grams are also represented by 32 bit hash values\footnote{If the $q$-gram length is larger than 16, some $q$-grams are mapped to the same hash value. However, we found out that longer words (e.g. 64 bit) only increases the memory consumption, but the mapping quality stays almost the same.}
The signature value is calculated using the \emph{exclusive or} operation (XOR) between each $q$-gram hash and the permutation value $\pi$ and then taking the minimum:

\begin{equation}
h_\pi(Q) = \min \{x \oplus \pi \;|\; x \in Q \}.
\label{eq:xor}
\end{equation}

Using 32 random bits and the XOR technique instead of a true random permutation means that the pre-conditions of Lemma~\ref{lemma:minhash} do not hold and the min-hash property may be violated \cite{Broder1998}.
However, empirical studies have shown that in practice the XOR technique approximates the desired property well \cite{Xor2013}.

Mapping each window to a single signature value is not enough to find a read. Therefore $s$ different permutation values are used. 
The parameter $s$ is called \emph{signature length}.

\subsection{Data structure}

For each permutation value a data structure, such as a simple hash table, is needed to map the calculated signature values (of each window) to the particular window in the reference genome.
Instead of a simple hash table, we use a two layer succinct rank data structure, because it needs less memory. 

An one layer rank data structure consists of a bit array $B$, an offset array $C$ and a data array $D$. 
The bit array is divided into blocks of length $\lambda$. 
For each block there is one entry in the offset array that indicates the number of 1's in $B$ up to this block.  
The data array contains one data entry for each 1 in $B$, so the length of $D$ is equal to the number of 1's in $B$. 
Given an index $i$ in $B$ with $B[i] = 1$, the corresponding data entry $D[j]$ can be efficiently accessed by using the offset array $C$.  
Additional information about rank data structures can be found for example in \cite{Navarro2016}.

This one layer rank data structure can be directly used for our purpose.
Each entry in $B$ represents a possible signature value, so the length of $B$ is $2^{32}$. 
The elements of the data array are also arrays that contain the particular window references.

This data structure has two disadvantages resulting in a very high memory consumption:
First, the array $B$ needs $\frac{2^{32}}{8}$ Bytes $= 512$ MB space. 
Note that this space is needed for each permutation value, so the memory consumption of $B$ must be multiplied with $s$, so for $s = 36$ we get a memory usage of already 18 GB for the bit arrays. 
Second, the most signature values are unique, i.e. the most arrays in $D$ contain only one window reference. 
This leads to a high memory overhead.

\begin{figure}
	\begin{center}
		\includegraphics[width=0.8\textwidth]{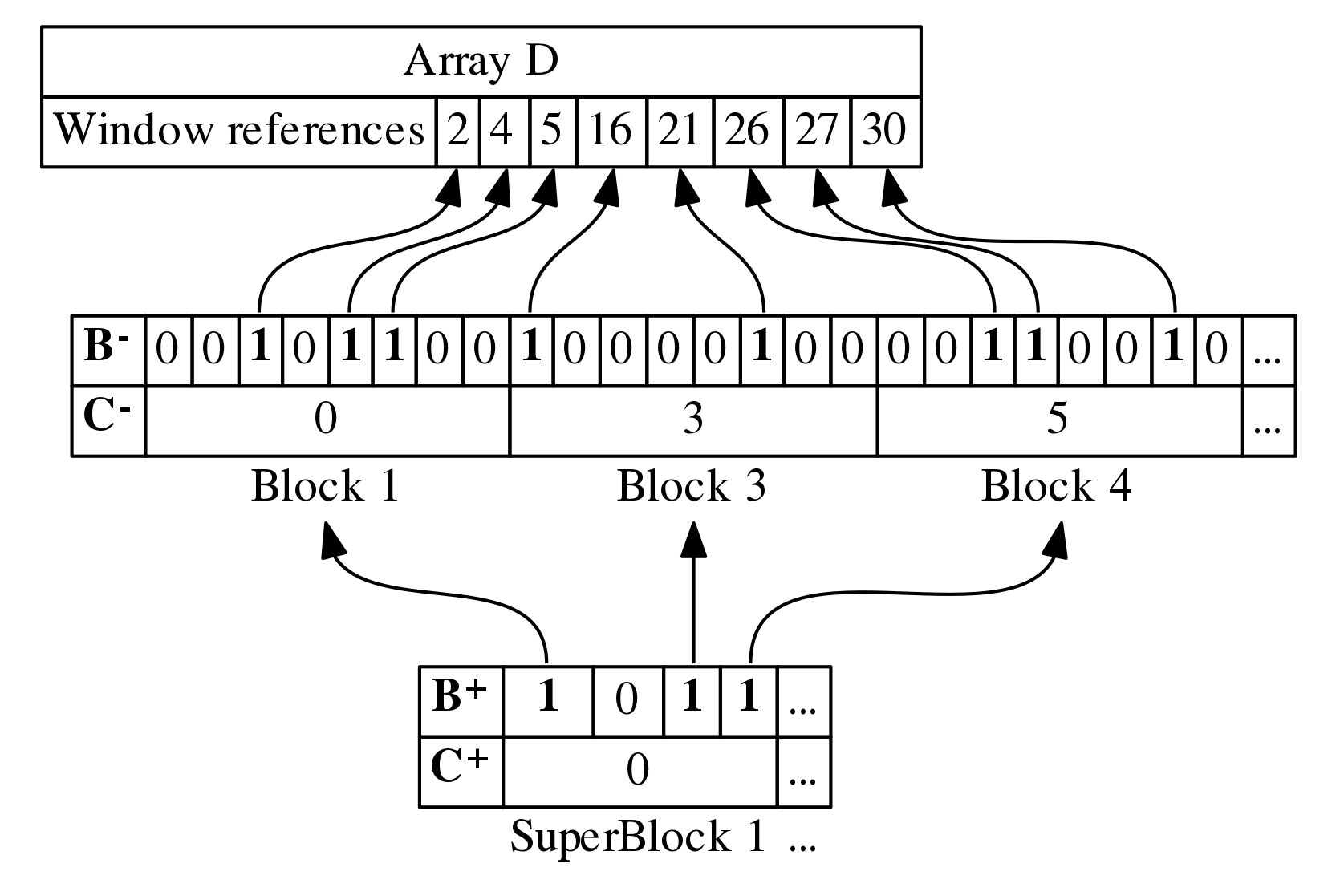}
	\end{center}
	\caption{Visualization of the two layer rank data structure. The figure was adapted from \cite{PG583}, some variable names were changes.}
	\label{fig:index:superrank}
\end{figure}

The first problem is solved by using a two layer rank data structure which is visualized in figure~\ref{fig:index:superrank}. For the human genome with 3 billion nucleotides, there are 24 million windows (when using the default window distance $o = 125$). 
Thus, only 0.6\%\footnote{Because $\frac{24\cdot10^6}{2^{32}} \approx 0.006$} of the entries are 1's and with $\lambda = 32$ at least 82\%\footnote{There are 24 million windows, so at most 24 million blocks can contain at least one ``1''. Thus, the ratio of blocks that contains only zeros is $1 - \frac{24 \cdot 10^6}{2^{32} / \lambda} = 0.8211...$} of the blocks contain only zeros. 
The two layer rank data structure uses the arrays $B^+$ and $C^+$ in the super layer and $B^-$, $C^-$ and $D^-$  in the second layer. 
Note that the data array $D^+$ of the super layer equals the arrays $B^-$ and $C^-$. 
The array $B^-$ contains only those blocks of $B$ that contain at least one ``1''.  
The super layer is needed to decide if a given block $k$ in $B$ is empty (i.e. $B^+[k] = 0$) or not (i.e. $B^+[k] = 1$). 
The data array $D^-$ is exactly the same as before, so $D^- = D$. By default, the block size of the super layer is $\lambda^+ = 64$ and the block size of the second layer is $\lambda^- = 32$. 

\begin{figure}
\begin{center}
\begin{tikzpicture}
	\tikzstyle{rect}=[fill=tudolight, thick]
	\tikzstyle{txt}=[inner sep=1pt, align=center]
	\tikzstyle{txtrect}=[draw, rectangle, rect, txt, inner sep=4pt]
	\tikzstyle{arrow}=[-latex, thick]

	\pgfmathsetmacro{\xWIEbegin}{0}
	\pgfmathsetmacro{\xWIEmid}{0.5}
	\pgfmathsetmacro{\xWIEbits}{(\xWIEbegin + \xWIEmid) / 2}
	\pgfmathsetmacro{\xWIEend}{3}
	\pgfmathsetmacro{\xWIEvalue}{(\xWIEmid + \xWIEend) / 2}
	\pgfmathsetmacro{\xWIEcenter}{(\xWIEbegin + \xWIEend) / 2}
	
	\pgfmathsetmacro{\yHead}{0.7}
	\pgfmathsetmacro{\yWIEsize}{0.5}
	\pgfmathsetmacro{\yWIEbeginA}{0}
	\pgfmathsetmacro{\yWIEendA}{\yWIEbeginA - \yWIEsize}
	\pgfmathsetmacro{\yWIEcenterA}{\yWIEbeginA - (\yWIEsize / 2)}
	\pgfmathsetmacro{\yWIEbeginB}{-1}
	\pgfmathsetmacro{\yWIEendB}{\yWIEbeginB - \yWIEsize}
	\pgfmathsetmacro{\yWIEcenterB}{\yWIEbeginB - (\yWIEsize / 2)}
	\pgfmathsetmacro{\yWIEbeginC}{-3}
	\pgfmathsetmacro{\yWIEendC}{\yWIEbeginC - \yWIEsize}
	\pgfmathsetmacro{\yWIEcenterC}{\yWIEbeginC - (\yWIEsize / 2)}
	\pgfmathsetmacro{\yWIEbeginD}{-5.9}
	\pgfmathsetmacro{\yWIEendD}{\yWIEbeginD - \yWIEsize}
	\pgfmathsetmacro{\yWIEcenterD}{\yWIEbeginD - (\yWIEsize / 2)}
	
	\pgfmathsetmacro{\xMWMbegin}{3.5}
	\pgfmathsetmacro{\xMWMend}{9.5}
	\pgfmathsetmacro{\xMWMcenter}{(\xMWMbegin + \xMWMend) / 2}
	\pgfmathsetmacro{\xMWMarrBegin}{4}
	\pgfmathsetmacro{\xMWMarrEnd}{9}
	\pgfmathsetmacro{\xMWMarrSize}{1}
	
	\pgfmathsetmacro{\yMWMarrBhead}{\yWIEbeginB}
	\pgfmathsetmacro{\yMWMarrBbegin}{\yMWMarrBhead - 0.5}
	\pgfmathsetmacro{\yMWMarrBend}{\yMWMarrBbegin - 1}
	\pgfmathsetmacro{\yMWMarrBmid}{(\yMWMarrBbegin + \yMWMarrBend)/2}
	\pgfmathsetmacro{\yMWMarrChead}{\yWIEbeginC}
	\pgfmathsetmacro{\yMWMarrCbegin}{\yMWMarrChead - 0.5}
	\pgfmathsetmacro{\yMWMarrCend}{\yMWMarrCbegin - 1.9}
	\pgfmathsetmacro{\yMWMarrCmid}{(\yMWMarrCbegin + \yMWMarrCend)/2}
	\pgfmathsetmacro{\yMWMarrDhead}{\yWIEbeginD}
	\pgfmathsetmacro{\yMWMarrDbegin}{\yMWMarrDhead - 0.5}
	\pgfmathsetmacro{\yMWMarrDend}{\yMWMarrDbegin - 0.5}
	\pgfmathsetmacro{\yMWMarrDextra}{\yMWMarrDend - 0.5}
	\pgfmathsetmacro{\yMWMarrDmid}{(\yMWMarrDbegin + \yMWMarrDend)/2}
	\pgfmathsetmacro{\yMWMrectBegin}{.25}
	\pgfmathsetmacro{\yMWMrectEnd}{\yMWMarrDextra - 3.5}
	
	\pgfmathsetmacro{\xOUTcenter}{11.2}
	
	\node[txt] at (\xWIEcenter, \yHead)
		{\textbf{Input} \\ \textit{Entry in $D^-$}};
	\node[txt] at (\xMWMcenter, \yHead)
		{\emph{\textbf{Multi Window Manager}}};
	\node[txt] at (\xOUTcenter, \yHead)
		{\textbf{Output} \\ \textit{Window references}};
	\draw[thick, dashed, fill=black!5] 
		(\xMWMbegin, \yMWMrectBegin) rectangle
		(\xMWMend, \yMWMrectEnd);
		
	\draw[rect] 
		(\xWIEbegin, \yWIEbeginA) rectangle
		(\xWIEmid, \yWIEendA);
	\draw[rect] 
		(\xWIEmid, \yWIEbeginA) rectangle
		(\xWIEend, \yWIEendA);
	\node[txt] at (\xWIEbits, \yWIEcenterA)
		{$00$};
	\node[txt] (Awie) at (\xWIEvalue, \yWIEcenterA)
		{$17$};
	\node[txt] (Aout) at (\xOUTcenter, \yWIEcenterA)
		{$\{17\}$};
	\draw[arrow] (Awie) -- (Aout);
		
	\draw[rect] 
		(\xWIEbegin, \yWIEbeginB) rectangle
		(\xWIEmid, \yWIEendB);
	\draw[rect] 
		(\xWIEmid, \yWIEbeginB) rectangle
		(\xWIEend, \yWIEendB);
	\node[txt] at (\xWIEbits, \yWIEcenterB)
		{$01$};
	\node[txt] (Bwie) at (\xWIEvalue, \yWIEcenterB)
		{$2$};
	\node[txt] (Bout) at (\xOUTcenter, \yWIEcenterB)
		{$\{4 , 8\}$};
		
	\node[txt] at (\xMWMcenter, \yMWMarrBhead) {\emph{doubleWindowBuckets}};
	\draw[rect] 
		(\xMWMarrBegin, \yMWMarrBbegin) rectangle
		(\xMWMarrEnd, \yMWMarrBend);
	\draw[rect] 
		(\xMWMarrBegin, \yMWMarrBbegin) rectangle
		(\xMWMarrBegin + \xMWMarrSize, \yMWMarrBend);
	\draw[rect] 
		(\xMWMarrBegin + \xMWMarrSize, \yMWMarrBbegin) rectangle
		(\xMWMarrBegin + \xMWMarrSize*2, \yMWMarrBend);
	\draw[rect] 
		(\xMWMarrBegin + \xMWMarrSize*2, \yMWMarrBbegin) rectangle
		(\xMWMarrBegin + \xMWMarrSize*3, \yMWMarrBend);
	\node[txt] at 
		(\xMWMarrBegin + \xMWMarrSize*0.5, \yMWMarrBmid) 
		{$3$\\[-2pt]$5$};
	\node[txt] at 
		(\xMWMarrBegin + \xMWMarrSize*1.5, \yMWMarrBmid) 
		{$26$\\[-2pt]$15$};
	\node[txt] (Bbucket) at
		(\xMWMarrBegin + \xMWMarrSize*2.5, \yMWMarrBmid)
		{$4$\\[-2pt]$8$};
	\node[txt] at 
		(\xMWMarrBegin + \xMWMarrSize*4, \yMWMarrBmid)
		{$\cdots$};
		
	\draw[arrow] (Bwie) -| (Bbucket.north west);
	\draw[arrow] (Bbucket.north east) |- (Bout);
		
	\draw[rect] 
		(\xWIEbegin, \yWIEbeginC) rectangle
		(\xWIEmid, \yWIEendC);
	\draw[rect] 
		(\xWIEmid, \yWIEbeginC) rectangle
		(\xWIEend, \yWIEendC);
	\node[txt] at (\xWIEbits, \yWIEcenterC)
		{$10$};
	\node[txt] (Cwie) at (\xWIEvalue, \yWIEcenterC)
		{$0$};
	\node[txt] (Cout) at (\xOUTcenter, \yWIEcenterC)
		{$\{4 , 8, 13\}$};
		
	\node[txt] at (\xMWMcenter, \yMWMarrChead) {\emph{quadrupleWindowBuckets}};
	\draw[rect] 
		(\xMWMarrBegin, \yMWMarrCbegin) rectangle
		(\xMWMarrEnd, \yMWMarrCend);
	\draw[rect] 
		(\xMWMarrBegin, \yMWMarrCbegin) rectangle
		(\xMWMarrBegin + \xMWMarrSize, \yMWMarrCend);
	\draw[rect] 
		(\xMWMarrBegin + \xMWMarrSize, \yMWMarrCbegin) rectangle
		(\xMWMarrBegin + \xMWMarrSize*2, \yMWMarrCend);
	\draw[rect] 
		(\xMWMarrBegin + \xMWMarrSize*2, \yMWMarrCbegin) rectangle
		(\xMWMarrBegin + \xMWMarrSize*3, \yMWMarrCend);
	\node[txt] (Cbucket) at 
		(\xMWMarrBegin + \xMWMarrSize*0.5, \yMWMarrCmid) 
		{$4$\\[-2pt]$8$\\[-2pt]$13$\\[-2pt]$-$};
	\node[txt]  at 
		(\xMWMarrBegin + \xMWMarrSize*1.5, \yMWMarrCmid) 
		{$6$\\[-2pt]$7$\\[-2pt]$23$\\[-2pt]$27$};
	\node[txt] at
		(\xMWMarrBegin + \xMWMarrSize*2.5, \yMWMarrCmid)
		{$16$\\[-2pt]$18$\\[-2pt]$28$\\[-2pt]$-$};
	\node[txt] at 
		(\xMWMarrBegin + \xMWMarrSize*4, \yMWMarrCmid)
		{$\cdots$};
		
	\draw[arrow] (Cwie) -| (Cbucket.north west);
	\draw[arrow] (Cbucket.north east) |- (Cout);
		
	\draw[rect] 
		(\xWIEbegin, \yWIEbeginD) rectangle
		(\xWIEmid, \yWIEendD);
	\draw[rect] 
		(\xWIEmid, \yWIEbeginD) rectangle
		(\xWIEend, \yWIEendD);
	\node[txt] at (\xWIEbits, \yWIEcenterD)
		{$11$};
	\node[txt] (Dwie) at (\xWIEvalue, \yWIEcenterD)
		{$1$};
	\node[txt] (Dout) at (\xOUTcenter, \yWIEcenterD)
		{$\{1,9,12,14,15\}$};
		
	\node[txt] at (\xMWMcenter, \yMWMarrDhead) {\emph{multiWindowBuckets}};
	\draw[rect] 
		(\xMWMarrBegin, \yMWMarrDbegin) rectangle
		(\xMWMarrEnd, \yMWMarrDend);
	\draw[rect] 
		(\xMWMarrBegin, \yMWMarrDbegin) rectangle
		(\xMWMarrBegin + \xMWMarrSize, \yMWMarrDend);
	\draw[rect] 
		(\xMWMarrBegin + \xMWMarrSize, \yMWMarrDbegin) rectangle
		(\xMWMarrBegin + \xMWMarrSize*2, \yMWMarrDend);
	\draw[rect] 
		(\xMWMarrBegin + \xMWMarrSize*2, \yMWMarrDbegin) rectangle
		(\xMWMarrBegin + \xMWMarrSize*3, \yMWMarrDend);
	\node[txt] (Dfirst) at 
		(\xMWMarrBegin + \xMWMarrSize*0.5, \yMWMarrDmid) 
		{};
	\node[txt] (Dbucket) at 
		(\xMWMarrBegin + \xMWMarrSize*1.5, \yMWMarrDmid) 
		{\ \ };
	\node[txt]  (Dthird) at
		(\xMWMarrBegin + \xMWMarrSize*2.5, \yMWMarrDmid)
		{};
	\node[txt] at 
		(\xMWMarrBegin + \xMWMarrSize*4, \yMWMarrDmid)
		{$\cdots$};
	\node[txtrect, below] (DfirstData) at 
		(\xMWMarrBegin + \xMWMarrSize*0.5, \yMWMarrDextra)
		{$2$\\[-2pt]$10$\\[-2pt]$11$\\[-2pt]$19$\\[-2pt]$21$\\[-2pt]$22$\\[-2pt]$25$};
	\node[txtrect, below] (DsecondData) at 
		(\xMWMarrBegin + \xMWMarrSize*1.5, \yMWMarrDextra)
		{$1$\\[-2pt]$9$\\[-2pt]$12$\\[-2pt]$14$\\[-2pt]$15$};
	\node[txt, below] (DthirdData) at 
		(\xMWMarrBegin + \xMWMarrSize*2.5, \yMWMarrDextra)
		{\rotatebox{90}{$\dots$}};
	\draw[arrow] (Dfirst) -- (DfirstData);
	\draw[arrow] (Dbucket) -- (DsecondData);
	\draw[arrow] (Dthird) -- (DthirdData);
	\draw[arrow] (Dwie) -| (Dbucket.north west);
	\draw[arrow] (Dbucket.north east) |- (Dout);
		
\end{tikzpicture}
\end{center}
\caption{Visualization of the \emph{MultiWindowManager}}
\label{fig:index:multiwindow}
\end{figure}
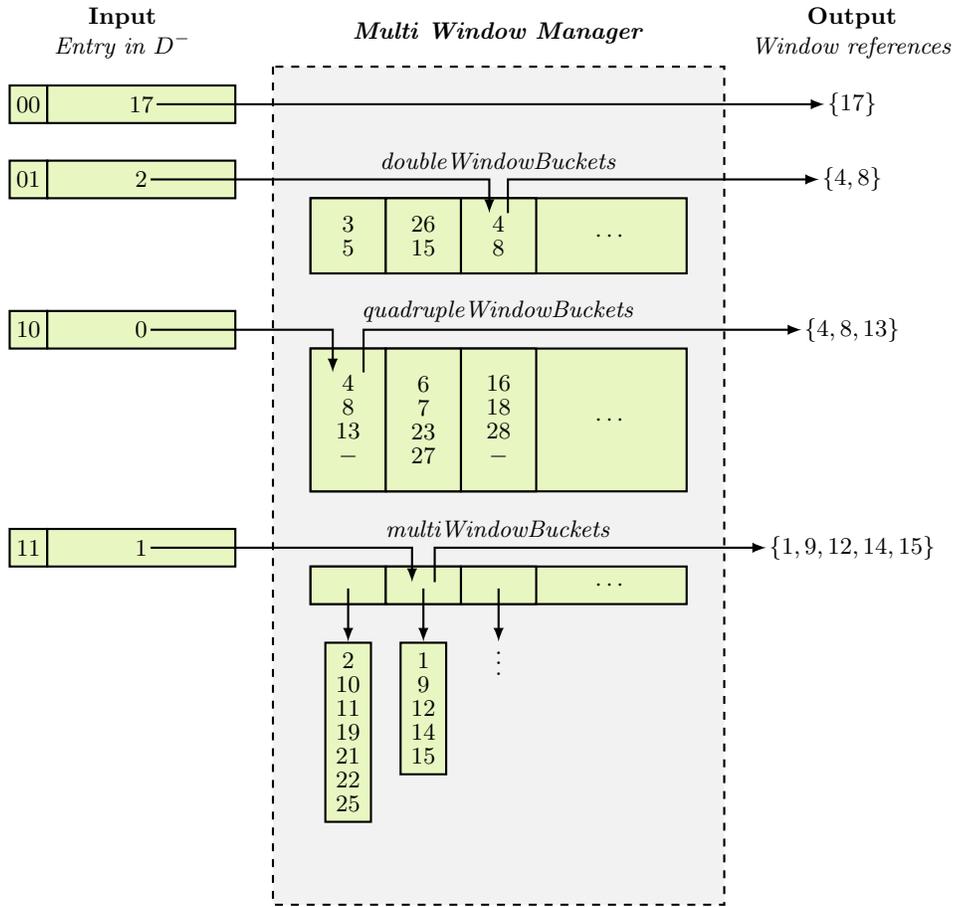

The second problem can be solved as follows. 
Usually a window reference is stored as a 32 bit value. 
However, for reasonable configurations there are always less than $2^{30}$ (about one billion) windows, so the two most significant bits can be used for meta information. 
If there is only one window for a given signature value, then the two bits are set to 00 and the window reference is stored directly in $D^-$. 
For signature values having exactly two corresponding windows, there is \emph{one} additional array whose elements consists of two window references. 
The corresponding entry in $D^-$ contains the index in that array and the meta bits are set to 01.
Furthermore there is one array whose elements consists of up to four window references and another one whose elements are dynamic arrays.
The described data structure is called \emph{Multi Window Manager} and visualized in figure~\ref{fig:index:multiwindow}.

Using more permutations leads to better results in the sense that we are better able to distinguish true similarities from random hits; however, the memory consumption increases linearly with the number of permutations.
For the human genome and reads of length $n = 100$ usually $s = 36$ permutation values are used which leads to a memory consumption of about 14 GB \cite{MA}.
For longer reads, the window distance relative to the read length ($o/n)$ can be decreased, so there are less windows and therefore less value pairs that have to be stored in the rank data structure, so the number of permutations can be increased.

\section{Read mapping}
\label{sec:read}

Mapping and aligning a DNA read is done in two phases.
In the mapping phase, candidate windows of the reference genome are determined and converted to candidate intervals.
In the alignment phase, a variant tolerant alignment between the read and the candidate intervals is performed to obtain the best alignment, taking all variants into account.

\subsection{Finding reference intervals}

To find a given read in the reference genome, first its $q$-grams set is determined.
Then the signature values are determined analogous to the index creation using the same permutation values.
Each signature value is looked up in the index data structure, so for each permutation value we receive a (possibly empty) list of window indices.
The same procedure is done with the reverse complement of the read.

Now we count how many times each window index was found.
This is done by sorting all found window indices.
To accelerate this, we sort each window list in the index data structure already during the index creation.
Then we use merge sort to merge the already sorted window lists.

If a read is located between two windows, there are common $q$-grams with both windows and therefore we probably obtain several common signature values for both windows.
For this reason neighboring windows are summarized to a so-called \emph{window sequence}.
Single windows (that have no neighbors) are also represented by a singleton window sequence.
Each window sequence is scored using the number of hits for each contained window. The score is calculated as follows.

Given a window sequence $\Omega$ that consists of $|\Omega|$ windows.
Let $c_1,\dots,c_{|\Omega|}$ be the number of hits for each contained window. 
If $|\Omega| = 1$, then the score is simply $c_1$. 
For longer window sequences the score $C(\Omega)$ is defined by
\begin{equation}
C(\Omega) \coloneqq \max_{i \in \{1,2,\dots,|\Omega|-1 \} } (c_i + c_{i+1})
\end{equation}

The idea is that a read cannot intersect with more than two windows if $n + w < 2o$ holds, which is fulfilled in the standard configuration.  Therefore, adding all $c_i$ of a long window sequence (i.e. $\sum_{i=1}^{|\Omega|} c_i$) would result in too large scores if there is a repetitive region in the reference region. 

The best scoring windows sequences are selected (by default at most $\kappa=64$).
Then each window sequence is converted to an interval on the reference genome.
Window sequences containing one window are enlarged depending on how often the window index was found (a window with only a few hits it is further enlarged than a window with many hits).
Window sequences containing two windows are contracted, because it is likely that the read is contained in both windows.
The more hits for the windows in the window sequence are found, the smaller is the resulting interval.

Let $\alpha > 0$ and $\beta_1 \leq 0$ be arbitrary constants, let $c_1$ be the number of hits for a singleton window sequence and $s$ the signature length. 
Then the window sequence is enlarged by 
\begin{equation}
w \cdot \alpha \left(1 - \frac{c_1}{s} \beta_1 \right)
\label{eqn:read:interval:enlarge}
\end{equation}
nucleotides on both sides, where $w$ is the window length. 
If $\beta_1 = 1$ and if we have a hit for each permutation, then $c1 = s$ holds and thus there is no enlargement.

Let $\beta_2$ be another arbitrary constant and $c_1$ and $c_2$ the number of hits for a window sequence of length 2. 
Let $a$ and $b$ be the start and end position of the window sequence (i.e. $a$ is the first position of the left window and $b$ is the last position of the right window). 
The resulting start position $a'$ and end position $b'$ of the interval are defined by:
\begin{align}
a' &\coloneqq a - n + q + \frac{c_2}{s} \cdot \beta_2 \cdot \frac{w - o + n - 2q}{2} 
\label{eqn:read:interval:contract:a} \\
b' &\coloneqq b + n - q - \frac{c_1}{s} \cdot \beta_2 \cdot \frac{w - o + n - 2q}{2}
\label{eqn:read:interval:contract:b}
\end{align}
If the number of hits are maximal and if $\beta_2 = 1$, then the resulting interval has the length $n$. 
The larger $c_1$ is, the more nucleotides of the read are probably located in the first window, so the right border of the interval can be contracted more than the left one.

If a window sequence consists of three or more windows, then the start and end position of the window sequence are used as interval.
By default the parameters have the values $\alpha = 0.43$ and $\beta_1 = \beta_2 = 0.3$.

\subsection{Variant tolerant alignment}

Each interval is processed by a variant tolerant aligner.
The aligner is based on Ukkonen's algorithm \cite{Ukkonen1985} and uses dynamic programming to calculate the optimal alignment according to the edit distance considering SNPs and indel variants.

Let $r = r_1, \ldots, r_m$ be the read and $t = t_1, \ldots, t_n$ be an interval on the reference genome.
To handle SNPs the characters of reference genome are not elements of the set $\Sigma = \{A,C,G,T\}$, but of its power set $\mathcal{P}(\Sigma)$. For example, $t_i = \{A,C\}$ means that there is a SNP variant at position $i$ and both nucleotides $A$ and $C$ co-exist at this position.

Let $F$ be a matrix of size $m \times n$. The element $F[i,j]$ denotes the edit distance  between $r_1 r_2 \cdots r_i$ and $t_{j'} t_{j'+1} \cdots t_j$ with some $j' \leq j$, so we have
\begin{align*}
F[i,0] =i &\quad  0 \leq i \leq m \\
F[0,j] =0 &\quad  1 \leq j \leq n
\end{align*}

The other matrix elements are calculated using the following recursive formula if there are no indel variants.
\begin{equation}
F[i,j] = \min \left\{
\begin{array}{ll} 
F[i-1,j-1] &+ \, [\![p_i \not\in t_j ]\!] \\ 
F[i-1,j] &+ \, 1 \\ 
F[i,j-1] &+ \, 1 
\end{array}\right.
\label{eq:align:recursive}
\end{equation}

Deletions can be considered by jumping back in the matrix.
If there is a deletion variant of length $k$ that ends at position $j$, such that the bases $t_{j-k},\dots, t_{j-1}$ are skipped, than the recursive formula for the column $j$ is  
\begin{equation}
F[i,j] = \min \left\{
\begin{array}{ll} 
F[i-1,j-1] &+ \, [\![p_i \not\in t_j ]\!] \\ 
F[i-1,j] &+ \, 1 \\ 
F[i,j-1] &+ \, 1 \\
F[i-1,j-k-1] &+ \, [\![p_i \not\in t_j ]\!] \\ 
F[i,j-k-1] &+ \, 1
\end{array}\right.
\label{eq:align:deletion}
\end{equation}
If there are more deletions that ends at positions~$j$, they can be added to the recursion formula by appending two more terms to the minimum expression.
So the recursion to be used at each reference position~$j$ is determined by the number and length(s) of the deletion variants ending at position~$j$. 

Handling insertions is slightly more complex.
If there is an insertion $s_1 s_2\cdots s_k$ before position $j$ (such that a string containing this insertion is for example $t_{j-1} s_1 \cdots s_k t_j$), then we need $k$ optional extra columns in the matrix $F$.
To calculate the column for $s_1$ we have to access the column for $t_{j-1}$.
The columns $s_2,\cdots,s_k$ can be determined straight-forwardly using the above recursion (\ref{eq:align:recursive}).
To calculate the column $t_j$ we not only have to access the column $t_{j-1}$, but also $s_k$ (analogous to equation~\eqref{eq:align:deletion}).
So for each insertion that ends before position $j$ two terms are added to the minimum expression. 

In contrast to the index of VATRAM, the aligner uses all available variants.

Filling the whole matrix consumes too much time.
Therefore the user can set an error threshold $k$, so that only the parts of the matrix whose values are less or equal to this threshold are calculated.
The details of this technique are more complicated, since there can be insertion or deletion variants and you have to ensure that all matrix elements that are accessed are already calculated.

The basic idea is to store for each matrix element an additional information called \emph{next row} which indicates the row index of the next element in the given column that is not larger than $k$. 
The \emph{next row} information of column $j$ is written during the calculation of the column $j-1$ in $F$.
Before an uninitialized element in $F$ will be accessed, it is set to $k+1$ beforehand. 
It can be shown that it does not matter that the correct value of that element is maybe larger than $k+1$.
In \cite{PG583} the details of this acceleration method are extensively described and the correctness of this pruning rule is proven.

The alignment is done for each given interval.
If the user is only interested in the best mapping, then the alignment cost of the first alignment can be used as error threshold for the second alignment to speed up the alignment process.

\subsection{Paired-end reads and long reads}

VATRAM is also able to map paired-end reads.
For that purpose both read sequences $u$ and $v$ of the paired-end read are first processed separately.
After creating the list of window sequences $L_u$ and $L_v$ for both read sequences, we look for window sequence pairs $(l_u, l_v) \in L_u \times L_v$, such that the distance between $l_u$ and $l_v$ fits to the insert size distribution of the paired-end reads.
To rank the window sequence pairs, the scoring values are combined to a single value. Finally, the alignment is done separately for both read sequences.

So far, we assume that all reads have a constant (or nearly constant) length~$n$.
However, some sequencing machines produce reads which length varies considerably.
For example PacBio reads \cite{PacBio2009} have a length from a few hundred nucleotides up to $60\,000$.
The window based approach of VATRAM needs reads of more or less constant length.
Thus, variably long reads are split into fragments whose length $\tilde{n}$ fits to the window length~$w$.
Then analogous to paired-end reads for each fragment the corresponding window sequence lists $L_1,\dots,L_k$ are calculated.
Afterwards, we look for tuples $(l_1, \dots, l_k) \in L_1 \times \dots \times L_k$ of windows sequences whose distances fits to the distance of the read fragments.
The scores of the window sequences are combined to a single value which is used for ranking the window sequence tuples.
If some fragments at the end or maybe in the mid of the read are not found, then there are some gaps in the tuple, so we allow that each $l_i$ is equal to an empty window sequence (or more formally $(l_1, \dots, l_k) \in L'_1 \times \dots \times L'_k$ where $L'_i \coloneqq L_i \cup \{\emptyset\}$ for all $1 \leq i \leq k$).
In this case the tuple gets a lower score, but the read can still be mapped if the tuple represents the correct position.
After ranking, each tuple is converted into an interval which is processed by the aligner.
The splitting procedure is applied if the fraction $n/w$ is larger than a given constant $f$ (by default $f = 0.8$ is used).

\section{Experiments}
\label{sec:exp}

We compare our read mapper VATRAM to other read mappers.
These are BWA-SW \cite{BWA2010}, as Bowtie2 \cite{Bowtie2012}, BWA-MEM \cite{BWA2013} in standard configuration and BWA-MEM configured such that it uses the edit distance as metric for the alignment (denoted as ``BWA-MEM in edit-distance configuraiton'').
Moreover we have tested \emph{mrsFastUltra} \cite{mrsFastUltra2014}, which is a variant tolerant read mapper like VATRAM.
Our read mapper was executed with and without using the known variants to quantify the benefit of using variants. 

VATRAMs parameter configuration is shown in table~\ref{tab:exp:config}. The formula for the signature length is an empirical result that leads to an approximately constant memory usage of about 14 GB independent of the window distance respectively read length. For $n = 100$ the signature length is $s = 36$. 

\begin{table}\centering
	\caption{Parameters used for the experiments. The splitting procedure was only used for data sets whose reads have variable length. In this case the parameters are $w = 140$, $o = 125$ and $s = 36$. The lower part of the table shows parameters that are only needed for read mapping and not for index creation (in contrast to the parameters in the upper part).}
	\label{tab:exp:config}
	\begin{tabular}{|l|l|r|}
		\hline
		\textbf{Symbol} & \textbf{Meaning} & \textbf{Value} \\
		\hline
		$w$ & window length & 1.4n \\
		$o$ & window distance & 1.25n \\
		$q$ & $q$-gram length & 17 \\
		$s$ & signature length & $\left[\frac{8.63n}{22.6 + 0.0138n}\right]$ \\
		$\delta$ & variant consideration threshold & 20\% \\
		$l$ & variant combination limit & 3 \\
		\hline
		$\kappa$ & maximal number of selected window sequences & 64 \\
		$\alpha$ & interval calculation & 0.43 \\
		$\beta_1$ &  interval calculation & 0.3 \\
		$\beta_2$ &  interval calculation & 0.3 \\
		$\tilde{n}$ & fragment length for reads with variable length & 100 \\
		$f$ & a read is split if $n/w > f$ & 0.8 \\
		\hline
	\end{tabular}
\end{table}

\subsection{Simulated reads}
\label{sec:exp:syn}

In our first experiment we used simulated single-end reads of constant length to compare VATRAM with other read mappers.
Using simulated reads has the advantage over real reads that the correct position is known. 

\begin{figure}[p]
\pgfplotsset{footnotesize,width=4.4cm,height=4.00cm} 
\pgfplotsset{mystyle/.style={thick, smooth, mark size = 3pt}}
\pgfplotsset{styleBWA/.style={mystyle, color=red, mark=x}}
\pgfplotsset{styleBWAedit/.style={mystyle, color=orange, mark=+}}
\pgfplotsset{styleBWAsw/.style={mystyle, color=orange!60!black, mark=asterisk}}
\pgfplotsset{styleBowtie/.style={mystyle, color=blue, mark=star}}
\pgfplotsset{styleVATRAM/.style={mystyle, color=green, mark=Mercedes star}}
\pgfplotsset{styleVATRAMnovar/.style={mystyle, color=green!75!cyan!50!black, mark=Mercedes star flipped}}
\pgfplotsset{styleMRSfast/.style={mystyle, color=cyan, mark=|}}
\begin{center}
\setlength{\tabcolsep}{-.5mm}
\begin{tabular}{rrr}
\begin{tikzpicture}
\begin{axis}[
mark options={solid},
ymax=1,ymin=0,
title={$f = 2\%$},
ylabel style={align=center},
ylabel=mapped correctly,
yticklabel= 
	\pgfmathparse{100*\tick}
	\pgfmathprintnumber{\pgfmathresult}\,\%,
yticklabel style={/pgf/number format/.cd,fixed,precision=2},
xtick={100,200,...,500},
legend columns=4,
legend entries={BWA-MEM, BWA-MEM (edit-distance), BWA-SW, Bowtie2, VATRAM, VATRAM (without variants), mrsFast-Ultra},
legend cell align=left,
legend to name=leg:exp:comp:const:rm
]
\addplot[styleBWA]
 table[x=rlng,y=bwa002] {data/exp_comp_const_syn_corr.dat};
\addplot[styleBWAedit]
 table[x=rlng,y=bwaedit002] {data/exp_comp_const_syn_corr.dat};
\addplot[styleBWAsw]
 table[x=rlng,y=bwasw002] {data/exp_comp_const_syn_corr.dat};
\addplot[styleBowtie]
 table[x=rlng,y=bowtie2002] {data/exp_comp_const_syn_corr.dat};
\addplot[styleVATRAM]
 table[x=rlng,y=vatram002] {data/exp_comp_const_syn_corr.dat};
\addplot[styleVATRAMnovar]
 table[x=rlng,y=vatramnovar002] {data/exp_comp_const_syn_corr.dat};
\addplot[styleMRSfast]
 table[x=rlng,y=mrsfast002] {data/exp_comp_const_syn_corr.dat};
\end{axis}
\end{tikzpicture}
&
\begin{tikzpicture}
\begin{axis}[
mark options={solid},
ymax=1,ymin=0,
title={$f = 4\%$},
ylabel style={align=center},
yticklabel=
	\pgfmathparse{100*\tick}
	\pgfmathprintnumber{\pgfmathresult}\,\%,
yticklabel style={/pgf/number format/.cd,fixed,precision=2},
xtick={100,200,...,500},
]
\addplot[styleBWA]
 table[x=rlng,y=bwa004] {data/exp_comp_const_syn_corr.dat};
\addplot[styleBWAedit]
 table[x=rlng,y=bwaedit004] {data/exp_comp_const_syn_corr.dat};
\addplot[styleBWAsw]
 table[x=rlng,y=bwasw004] {data/exp_comp_const_syn_corr.dat};
\addplot[styleBowtie]
 table[x=rlng,y=bowtie2004] {data/exp_comp_const_syn_corr.dat};
\addplot[styleVATRAM]
 table[x=rlng,y=vatram004] {data/exp_comp_const_syn_corr.dat};
\addplot[styleVATRAMnovar]
 table[x=rlng,y=vatramnovar004] {data/exp_comp_const_syn_corr.dat};
\addplot[styleMRSfast]
 table[x=rlng,y=mrsfast004] {data/exp_comp_const_syn_corr.dat};
\end{axis}
\end{tikzpicture}
&
\begin{tikzpicture}
\begin{axis}[
mark options={solid},
ymax=1,ymin=0,
title={$f = 8\%$},
ylabel style={align=center},
yticklabel=
	\pgfmathparse{100*\tick}
	\pgfmathprintnumber{\pgfmathresult}\,\%,
yticklabel style={/pgf/number format/.cd,fixed,precision=2},
xtick={100,200,...,500},
]
\addplot[styleBWA]
 table[x=rlng,y=bwa008] {data/exp_comp_const_syn_corr.dat};
\addplot[styleBWAedit]
 table[x=rlng,y=bwaedit008] {data/exp_comp_const_syn_corr.dat};
\addplot[styleBWAsw]
 table[x=rlng,y=bwasw008] {data/exp_comp_const_syn_corr.dat};
\addplot[styleBowtie]
 table[x=rlng,y=bowtie2008] {data/exp_comp_const_syn_corr.dat};
\addplot[styleVATRAM]
 table[x=rlng,y=vatram008] {data/exp_comp_const_syn_corr.dat};
\addplot[styleVATRAMnovar]
 table[x=rlng,y=vatramnovar008] {data/exp_comp_const_syn_corr.dat};
\addplot[styleMRSfast]
 table[x=rlng,y=mrsfast008] {data/exp_comp_const_syn_corr.dat};
\end{axis}
\end{tikzpicture}
\\
\begin{tikzpicture}
\begin{axis}[
mark options={solid},
ymax=1,ymin=0.94,
ylabel style={align=center},
ylabel=mapped correctly\\ (details),
yticklabel=
	\pgfmathparse{100*\tick}
	\pgfmathprintnumber{\pgfmathresult}\,\%,
yticklabel style={/pgf/number format/.cd,fixed,precision=2},
xtick={100,200,...,500},
]
\addplot[styleBWA]
 table[x=rlng,y=bwa002] {data/exp_comp_const_syn_corr.dat};
\addplot[styleBWAedit]
 table[x=rlng,y=bwaedit002] {data/exp_comp_const_syn_corr.dat};
\addplot[styleBWAsw]
 table[x=rlng,y=bwasw002] {data/exp_comp_const_syn_corr.dat};
\addplot[styleBowtie]
 table[x=rlng,y=bowtie2002] {data/exp_comp_const_syn_corr.dat};
\addplot[styleVATRAM]
 table[x=rlng,y=vatram002] {data/exp_comp_const_syn_corr.dat};
\addplot[styleVATRAMnovar]
 table[x=rlng,y=vatramnovar002] {data/exp_comp_const_syn_corr.dat};
\addplot[styleMRSfast]
 table[x=rlng,y=mrsfast002] {data/exp_comp_const_syn_corr.dat};
\end{axis}
\end{tikzpicture}
&
\begin{tikzpicture}
\begin{axis}[
mark options={solid},
ymax=1,ymin=0.9,
ylabel style={align=center},
yticklabel=
	\pgfmathparse{100*\tick}
	\pgfmathprintnumber{\pgfmathresult}\,\%,
yticklabel style={/pgf/number format/.cd,fixed,precision=2},
xtick={100,200,...,500},
]
\addplot[styleBWA]
 table[x=rlng,y=bwa004] {data/exp_comp_const_syn_corr.dat};
\addplot[styleBWAedit]
 table[x=rlng,y=bwaedit004] {data/exp_comp_const_syn_corr.dat};
\addplot[styleBWAsw]
 table[x=rlng,y=bwasw004] {data/exp_comp_const_syn_corr.dat};
\addplot[styleBowtie]
 table[x=rlng,y=bowtie2004] {data/exp_comp_const_syn_corr.dat};
\addplot[styleVATRAM]
 table[x=rlng,y=vatram004] {data/exp_comp_const_syn_corr.dat};
\addplot[styleVATRAMnovar]
 table[x=rlng,y=vatramnovar004] {data/exp_comp_const_syn_corr.dat};
\addplot[styleMRSfast]
 table[x=rlng,y=mrsfast004] {data/exp_comp_const_syn_corr.dat};
\end{axis}
\end{tikzpicture}
&
\begin{tikzpicture}
\begin{axis}[
mark options={solid},
ymax=1,ymin=0.75,
ylabel style={align=center},
yticklabel=
	\pgfmathparse{100*\tick}
	\pgfmathprintnumber{\pgfmathresult}\,\%,
yticklabel style={/pgf/number format/.cd,fixed,precision=2},
xtick={100,200,...,500},
]
\addplot[styleBWA]
 table[x=rlng,y=bwa008] {data/exp_comp_const_syn_corr.dat};
\addplot[styleBWAedit]
 table[x=rlng,y=bwaedit008] {data/exp_comp_const_syn_corr.dat};
\addplot[styleBWAsw]
 table[x=rlng,y=bwasw008] {data/exp_comp_const_syn_corr.dat};
\addplot[styleBowtie]
 table[x=rlng,y=bowtie2008] {data/exp_comp_const_syn_corr.dat};
\addplot[styleVATRAM]
 table[x=rlng,y=vatram008] {data/exp_comp_const_syn_corr.dat};
\addplot[styleVATRAMnovar]
 table[x=rlng,y=vatramnovar008] {data/exp_comp_const_syn_corr.dat};
\addplot[styleMRSfast]
 table[x=rlng,y=mrsfast008] {data/exp_comp_const_syn_corr.dat};
\end{axis}
\end{tikzpicture}
\\
\begin{tikzpicture}
\begin{axis}[
mark options={solid},
ymin=0,
ylabel style={align=center},
ylabel={mapped wrongly},
yticklabel=
	\pgfmathparse{100*\tick}
	\pgfmathprintnumber{\pgfmathresult}\,\%,
scaled ticks=false,
xtick={100,200,...,500},
]
\addplot[styleBWA]
 table[x=rlng,y=bwa002] {data/exp_comp_const_syn_wrong.dat};
\addplot[styleBWAedit]
 table[x=rlng,y=bwaedit002] {data/exp_comp_const_syn_wrong.dat};
\addplot[styleBWAsw]
 table[x=rlng,y=bwasw002] {data/exp_comp_const_syn_wrong.dat};
\addplot[styleBowtie]
 table[x=rlng,y=bowtie2002] {data/exp_comp_const_syn_wrong.dat};
\addplot[styleVATRAM]
 table[x=rlng,y=vatram002] {data/exp_comp_const_syn_wrong.dat};
\addplot[styleVATRAMnovar]
 table[x=rlng,y=vatramnovar002] {data/exp_comp_const_syn_wrong.dat};
\addplot[styleMRSfast]
 table[x=rlng,y=mrsfast002] {data/exp_comp_const_syn_wrong.dat};
\end{axis}
\end{tikzpicture}
&
\begin{tikzpicture}
\begin{axis}[
mark options={solid},
ymin=0,
={$f = 4\%$},
ylabel style={align=center},
yticklabel=
	\pgfmathparse{100*\tick}
	\pgfmathprintnumber{\pgfmathresult}\,\%,
scaled ticks=false,
xtick={100,200,...,500},
]
\addplot[styleBWA]
 table[x=rlng,y=bwa004] {data/exp_comp_const_syn_wrong.dat};
\addplot[styleBWAedit]
 table[x=rlng,y=bwaedit004] {data/exp_comp_const_syn_wrong.dat};
\addplot[styleBWAsw]
 table[x=rlng,y=bwasw004] {data/exp_comp_const_syn_wrong.dat};
\addplot[styleBowtie]
 table[x=rlng,y=bowtie2004] {data/exp_comp_const_syn_wrong.dat};
\addplot[styleVATRAM]
 table[x=rlng,y=vatram004] {data/exp_comp_const_syn_wrong.dat};
\addplot[styleVATRAMnovar]
 table[x=rlng,y=vatramnovar004] {data/exp_comp_const_syn_wrong.dat};
\addplot[styleMRSfast]
 table[x=rlng,y=mrsfast004] {data/exp_comp_const_syn_wrong.dat};
\end{axis}
\end{tikzpicture}
&
\begin{tikzpicture}
\begin{axis}[
mark options={solid},
ymin=0,
ylabel style={align=center},
yticklabel=
	\pgfmathparse{100*\tick}
	\pgfmathprintnumber{\pgfmathresult}\,\%,
yticklabel style={/pgf/number format/.cd,fixed,precision=2},
scaled ticks=false,
xtick={100,200,...,500},
]
\addplot[styleBWA]
 table[x=rlng,y=bwa008] {data/exp_comp_const_syn_wrong.dat};
\addplot[styleBWAedit]
 table[x=rlng,y=bwaedit008] {data/exp_comp_const_syn_wrong.dat};
\addplot[styleBWAsw]
 table[x=rlng,y=bwasw008] {data/exp_comp_const_syn_wrong.dat};
\addplot[styleBowtie]
 table[x=rlng,y=bowtie2008] {data/exp_comp_const_syn_wrong.dat};
\addplot[styleVATRAM]
 table[x=rlng,y=vatram008] {data/exp_comp_const_syn_wrong.dat};
\addplot[styleVATRAMnovar]
 table[x=rlng,y=vatramnovar008] {data/exp_comp_const_syn_wrong.dat};
\addplot[styleMRSfast]
 table[x=rlng,y=mrsfast008] {data/exp_comp_const_syn_wrong.dat};
\end{axis}
\end{tikzpicture}
\\
\begin{tikzpicture}
\begin{axis}[
mark options={solid},
ymin=0,
ylabel style={align=center},
ylabel=runtime per\\ nucleotide (in \textmu s),
yticklabel style={/pgf/number format/.cd,fixed,precision=2},
xtick={100,200,...,500},
]
\addplot[styleBWA]
 table[x=rlng,y=bwa002] {data/exp_comp_const_syn_tbase.dat};
\addplot[styleBWAedit]
 table[x=rlng,y=bwaedit002] {data/exp_comp_const_syn_tbase.dat};
\addplot[styleBWAsw]
 table[x=rlng,y=bwasw002] {data/exp_comp_const_syn_tbase.dat};
\addplot[styleBowtie]
 table[x=rlng,y=bowtie2002] {data/exp_comp_const_syn_tbase.dat};
\addplot[styleVATRAM]
 table[x=rlng,y=vatram002] {data/exp_comp_const_syn_tbase.dat};
\addplot[styleVATRAMnovar]
 table[x=rlng,y=vatramnovar002] {data/exp_comp_const_syn_tbase.dat};
\addplot[styleMRSfast]
 table[x=rlng,y=mrsfast002] {data/exp_comp_const_syn_tbase.dat};
\end{axis}
\end{tikzpicture}
&
\begin{tikzpicture}
\begin{axis}[
mark options={solid},
ymin=0,
ylabel style={align=center},
yticklabel style={/pgf/number format/.cd,fixed,precision=2},
xtick={100,200,...,500},
]
\addplot[styleBWA]
 table[x=rlng,y=bwa004] {data/exp_comp_const_syn_tbase.dat};
\addplot[styleBWAedit]
 table[x=rlng,y=bwaedit004] {data/exp_comp_const_syn_tbase.dat};
\addplot[styleBWAsw]
 table[x=rlng,y=bwasw004] {data/exp_comp_const_syn_tbase.dat};
\addplot[styleBowtie]
 table[x=rlng,y=bowtie2004] {data/exp_comp_const_syn_tbase.dat};
\addplot[styleVATRAM]
 table[x=rlng,y=vatram004] {data/exp_comp_const_syn_tbase.dat};
\addplot[styleVATRAMnovar]
 table[x=rlng,y=vatramnovar004] {data/exp_comp_const_syn_tbase.dat};
\addplot[styleMRSfast]
 table[x=rlng,y=mrsfast004] {data/exp_comp_const_syn_tbase.dat};
\end{axis}
\end{tikzpicture}
&
\begin{tikzpicture}
\begin{axis}[
mark options={solid},
ymin=0,
ylabel style={align=center},
yticklabel style={/pgf/number format/.cd,fixed,precision=2},
xtick={100,200,...,500},
]
\addplot[styleBWA]
 table[x=rlng,y=bwa008] {data/exp_comp_const_syn_tbase.dat};
\addplot[styleBWAedit]
 table[x=rlng,y=bwaedit008] {data/exp_comp_const_syn_tbase.dat};
\addplot[styleBWAsw]
 table[x=rlng,y=bwasw008] {data/exp_comp_const_syn_tbase.dat};
\addplot[styleBowtie]
 table[x=rlng,y=bowtie2008] {data/exp_comp_const_syn_tbase.dat};
\addplot[styleVATRAM]
 table[x=rlng,y=vatram008] {data/exp_comp_const_syn_tbase.dat};
\addplot[styleVATRAMnovar]
 table[x=rlng,y=vatramnovar008] {data/exp_comp_const_syn_tbase.dat};
\addplot[styleMRSfast]
 table[x=rlng,y=mrsfast008] {data/exp_comp_const_syn_tbase.dat};
\end{axis}
\end{tikzpicture}
\\
\begin{tikzpicture}
\begin{axis}[
mark options={solid},
ymin=0,
xlabel=read length,
ylabel style={align=center},
ylabel=memory \\ consumption \\ (in GB),
yticklabel style={/pgf/number format/.cd,fixed,precision=2},
xtick={100,200,...,500},
]
\addplot[styleBWA]
 table[x=rlng,y=bwa002] {data/exp_comp_const_syn_mem.dat};
\addplot[styleBWAedit]
 table[x=rlng,y=bwaedit002] {data/exp_comp_const_syn_mem.dat};
\addplot[styleBWAsw]
 table[x=rlng,y=bwasw002] {data/exp_comp_const_syn_mem.dat};
\addplot[styleBowtie]
 table[x=rlng,y=bowtie2002] {data/exp_comp_const_syn_mem.dat};
\addplot[styleVATRAM]
 table[x=rlng,y=vatram002] {data/exp_comp_const_syn_mem.dat};
\addplot[styleVATRAMnovar]
 table[x=rlng,y=vatramnovar002] {data/exp_comp_const_syn_mem.dat};
\addplot[styleMRSfast]
 table[x=rlng,y=mrsfast002] {data/exp_comp_const_syn_mem.dat};
\end{axis}
\end{tikzpicture}
&
\begin{tikzpicture}
\begin{axis}[
mark options={solid},
ymin=0,
xlabel=read length,
ylabel style={align=center},
yticklabel style={/pgf/number format/.cd,fixed,precision=2},
xtick={100,200,...,500},
]
\addplot[styleBWA]
 table[x=rlng,y=bwa004] {data/exp_comp_const_syn_mem.dat};
\addplot[styleBWAedit]
 table[x=rlng,y=bwaedit004] {data/exp_comp_const_syn_mem.dat};
\addplot[styleBWAsw]
 table[x=rlng,y=bwasw004] {data/exp_comp_const_syn_mem.dat};
\addplot[styleBowtie]
 table[x=rlng,y=bowtie2004] {data/exp_comp_const_syn_mem.dat};
\addplot[styleVATRAM]
 table[x=rlng,y=vatram004] {data/exp_comp_const_syn_mem.dat};
\addplot[styleVATRAMnovar]
 table[x=rlng,y=vatramnovar004] {data/exp_comp_const_syn_mem.dat};
\addplot[styleMRSfast]
 table[x=rlng,y=mrsfast004] {data/exp_comp_const_syn_mem.dat};
\end{axis}
\end{tikzpicture}
&
\begin{tikzpicture}
\begin{axis}[
mark options={solid},
ymin=0,
xlabel=read length,
ylabel style={align=center},
yticklabel style={/pgf/number format/.cd,fixed,precision=2},
xtick={100,200,...,500},
]
\addplot[styleBWA]
 table[x=rlng,y=bwa008] {data/exp_comp_const_syn_mem.dat};
\addplot[styleBWAedit]
 table[x=rlng,y=bwaedit008] {data/exp_comp_const_syn_mem.dat};
\addplot[styleBWAsw]
 table[x=rlng,y=bwasw008] {data/exp_comp_const_syn_mem.dat};
\addplot[styleBowtie]
 table[x=rlng,y=bowtie2008] {data/exp_comp_const_syn_mem.dat};
\addplot[styleVATRAM]
 table[x=rlng,y=vatram008] {data/exp_comp_const_syn_mem.dat};
\addplot[styleVATRAMnovar]
 table[x=rlng,y=vatramnovar008] {data/exp_comp_const_syn_mem.dat};
\addplot[styleMRSfast]
 table[x=rlng,y=mrsfast008] {data/exp_comp_const_syn_mem.dat};
\end{axis}
\end{tikzpicture}
\end{tabular}
\ref*{leg:exp:comp:const:rm}
\end{center}
\caption{Comparison between different readmappers using simulated reads. The error rate was 2\%, 4\% and 8\% per position (the error rate in the diagrams of one column is constant and shown above the diagrams of the first row). The x-axis shows the read length. The ratio of correctly mapped reads is shown twice (first and second row) to improve the visibility.}
\label{fig:exp:syn:se}
\end{figure}
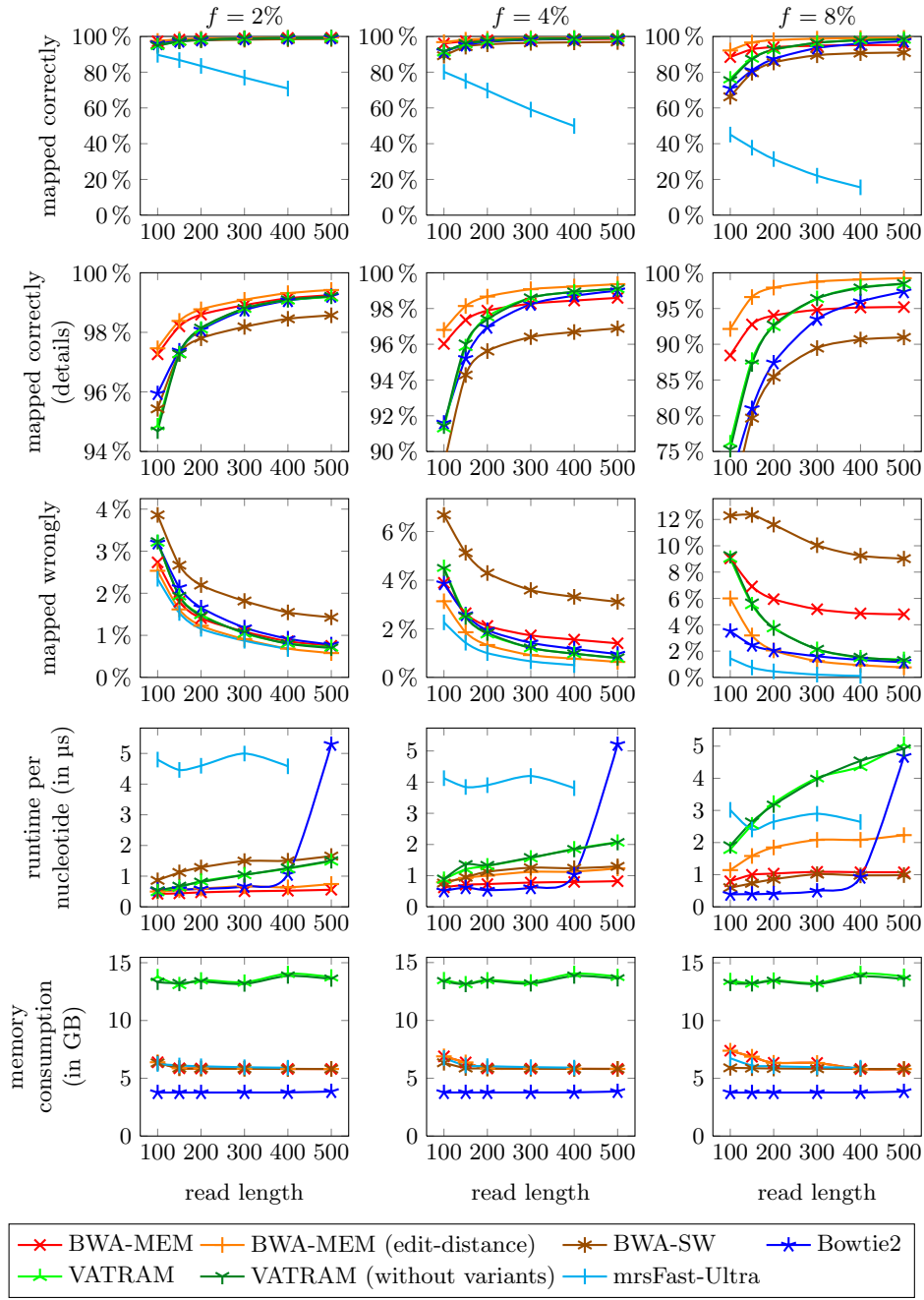

The reads were created from the human reference genome (GRCh37).
Each known variant is inserted into the reads with its population frequency.
Furthermore sequencing errors respectively unknown variants are inserted in the reads with a probability of 2\%, 4\% or 8\% per position.
Of the errors, 90\% are substitutions of single nucleotides, the remaining 10\% are insertions or deletions of variable length.
The ratio of the correct and wrong mapped reads as well as the runtime and the memory consumption are shown in figure~\ref{fig:exp:syn:se} for different read lengths.
Note that the sum of the correct and wrong mapped reads is not equal to 100\%, because there can be reads that remain unmapped by the particular read mapped, so the sum is equal to the fraction of the reads that were mapped.

As we can see the memory consumption of VATRAM (about 14~GB) is significantly higher than those of the other read mappers.
As most modern workstations have 16~GB of memory or even more, we do not expect this to be a major disadvantage.

The runtime of VATRAM is (in most cases) higher than the runtime of BWA or Bowtie2.
The reason for this is that VATRAM takes variants into account and thus solves a problem that is more complex than the alignment that is done by BWA or Bowtie2.
In fact, most of the runtime of VATRAM is needed to calculate the variant tolerant alignment.
The mapping process (i.e. calculating the signature values of a read, looking them up to get the window indices and converting them to intervals) is much faster (approximately five times) than the alignment calculation\cite{MA}.
The read mapper \emph{mrsFastUltra} needs more time than VATRAM if the error rate is less or equal to 4\%.
With an error rate of 8\%, \emph{mrsFastUltra} is faster than VATRAM, but then over 50\% of the reads are left unmapped.

VATRAM maps more reads to the correct position than \emph{mrsFastUltra} which is also variant tolerant.
However, BWA-MEM produces even better results (especially for the short reads) although it has no information about the known variants.
The number of wrongly mapped reads of VATRAM is small, similar to BWA-MEM and Bowtie2.
The lowest fraction of wrongly mapped reads is achieved by \emph{mrsFastUltra}.
However, this is not surprising, since \emph{mrsFastUltra} can only map a small number of reads, so many reads that are difficult to map correctly are left unmapped.

Although the simulated reads contain variants, there is no great benefit to consider them with VATRAM.
Only for a read length of $n = 100$ and an error of 2\%, the improvement is visible in figure~\ref{fig:exp:syn:se}.
However, the fraction of correctly mapped reads only increases by 0.18\%-points to 94.86\%. This is considerably fewer than the results of BWA-MEM in edit-distance configuration with 97.4\%.
The reason for this small improvement is that most known variants appears with a frequency of less than 20\%.
These variants are not considered by VATRAMs index, because the variant consideration threshold was $\delta = 0.2$.
A lower threshold would lead to even worse results \cite{MA}, since reads that do not contain the variant (these are more than 80\%) are found with a lower probability if the variant is added to the index.

In this experiment, VATRAM did not produce better results than BWA-MEM, although it uses extra knowledge about the known variants.
However, VATRAM outperformed the variant-toleratn read mapper \emph{mrsFastUltra}.
The benefit of considering the variants was very low.

\subsection{Real reads}
\label{sec:exp:real}

In this section we use real data sets to compare the read mappers.
When using real reads, the correct position is not known.
One solution is to compare only the number of mapped reads.
However, this is not a reasonable measure, because one may map a read to an arbitrary position which would lead to the best possible performance under this evaluation metric.
Another method is to compare the edit distance of the different alignments.
However, calculating the edit distance between a read and the corresponding interval on the reference genome does not consider any variants and therefore the unknown correct position of the read would not necessarily have the smallest distance. 

Therefore we decided to use VATRAM's aligner to realign the reads to the interval on the reference genome given by the different read mappers.
A mapping is defined as correct if there is no other mapping that leads to a lower edit distance considering all known variants.
If two or more different mappings have the same optimal alignment cost, then all are treated as correct.
Note that this does not provide VATRAM with an advantage because the aligner of VATRAM is just a tool to calculate the optimal alignment considering all known variants.

Table~\ref{tab:exp:real:overview} shows an overview of the different data sets.
There are two paired-end data sets with a constant read length as well as three single-end data sets with a variable read length created with three different sequencing machines.

\begin{table}[bth]\centering
\caption{Data sets used in the experiments.
The data set name refers to NCBI sequence read archive (SRA).
Column $n$ shows the average read length, column \emph{Type} indicates whether the data set contains paired-end (PE) ord single-end (SE) reads and the column \emph{Const?} shows if all reads in the data set have the same length (\emph{yes}) or not (\emph{no}).
The last column shows the error threshold that was used for the read mapper VATRAM.
Note that the later alignment that is executed for each read mapper is done with a much higher error threshold to ensure that always the best alignment is calculated.}
\label{tab:exp:real:overview}
\begin{tabular}{|l|l|r|l|l|r|l|}
\hline
Name & Sequencing machine & $n$ & Type & Const? & Err.thr. \\
\hline
ERR259389 & Illumina MiSeq & 151 & PE & yes & 15 \\
ERR967952 & Illumina MiSeq & 250 & PE & yes & 25 \\
\hline
DRR003760 & Illumina MiSeq & 180 & SE & no & 20 \\
SRR003174 & 454 GS FLX Titanium & 565 & SE & no & 60 \\
SRX533609 & PacBio RS II & 8651 & SE & no & 450 \\
\hline
\end{tabular}
\end{table}

\subsubsection{Real paired-end reads}
\label{sec:exp:real:pe}

\begin{figure}[tp]
\pgfplotsset{footnotesize,width=5cm,height=4.6cm}
\begin{center}
\setlength{\tabcolsep}{0mm}
\begin{tabular}{rr}
\begin{tikzpicture}
\begin{axis}[
ybar,
ymin=0.86,ymax=1,
enlarge x limits=0.2,
bar width=0.25cm,
ylabel style={align=center},
ylabel=mapped correctly,
yticklabel=
	\pgfmathparse{100*\tick}
	\pgfmathprintnumber{\pgfmathresult}\,\%,
yticklabel style={/pgf/number format/.cd,fixed,precision=2},
symbolic x coords={ERR259389},
xtick=data,
scaled ticks=false,
legend columns=4,
legend entries={BWA-MEM, BWA-MEM (edit-distance), BWA-SW, Bowtie2, VATRAM, VATRAM (without variants), mrsFast-Ultra},
legend cell align=left,
legend to name=leg:exp:rm:bars
]
\addplot[barstyleBWA]
 table[x=set,y=bwa] {data/exp_comp_pe_real_corr_b.dat};
\addplot[barstyleBWAedit]
 table[x=set,y=bwaedit] {data/exp_comp_pe_real_corr_b.dat};
\addplot[barstyleBWAsw]
 table[x=set,y=bwasw] {data/exp_comp_pe_real_corr_b.dat};
\addplot[barstyleBowtie]
 table[x=set,y=bowtie2] {data/exp_comp_pe_real_corr_b.dat};
\addplot[barstyleVATRAM]
 table[x=set,y=vatram] {data/exp_comp_pe_real_corr_b.dat};
\addplot[barstyleVATRAMnovar]
 table[x=set,y=vatramnovar] {data/exp_comp_pe_real_corr_b.dat};
\addplot[barstyleMRSfast]
 table[x=set,y=mrsfast] {data/exp_comp_pe_real_corr_b.dat};
\end{axis}
\end{tikzpicture}
&
\begin{tikzpicture}
\begin{axis}[
ybar,
ymin=0.0,ymax=1,
enlarge x limits=0.2,
bar width=0.25cm,
ylabel style={align=center},
yticklabel=
	\pgfmathparse{100*\tick}
	\pgfmathprintnumber{\pgfmathresult}\,\%,
yticklabel style={/pgf/number format/.cd,fixed,precision=2},
symbolic x coords={ERR967952},
xtick=data,
scaled ticks=false,
]
\addplot[barstyleBWA]
 table[x=set,y=bwa] {data/exp_comp_pe_real_corr_c.dat};
\addplot[barstyleBWAedit]
 table[x=set,y=bwaedit] {data/exp_comp_pe_real_corr_c.dat};
\addplot[barstyleBWAsw]
 table[x=set,y=bwasw] {data/exp_comp_pe_real_corr_c.dat};
\addplot[barstyleBowtie]
 table[x=set,y=bowtie2] {data/exp_comp_pe_real_corr_c.dat};
\addplot[barstyleVATRAM]
 table[x=set,y=vatram] {data/exp_comp_pe_real_corr_c.dat};
\addplot[barstyleVATRAMnovar]
 table[x=set,y=vatramnovar] {data/exp_comp_pe_real_corr_c.dat};
\addplot[barstyleMRSfast]
 table[x=set,y=mrsfast] {data/exp_comp_pe_real_corr_c.dat};
\end{axis}
\end{tikzpicture}
\\
\begin{tikzpicture}
\begin{axis}[
ybar,
ymin=0.0,
enlarge x limits=0.2,
bar width=0.25cm,
ylabel style={align=center},
ylabel=mapped wrongly,
yticklabel=
	\pgfmathparse{100*\tick}
	\pgfmathprintnumber{\pgfmathresult}\,\%,
yticklabel style={/pgf/number format/.cd,fixed,precision=2},
symbolic x coords={ERR259389},
xtick=data,
scaled ticks=false,
]
\addplot[barstyleBWA]
 table[x=set,y=bwa] {data/exp_comp_pe_real_wrong_b.dat};
\addplot[barstyleBWAedit]
 table[x=set,y=bwaedit] {data/exp_comp_pe_real_wrong_b.dat};
\addplot[barstyleBWAsw]
 table[x=set,y=bwasw] {data/exp_comp_pe_real_wrong_b.dat};
\addplot[barstyleBowtie]
 table[x=set,y=bowtie2] {data/exp_comp_pe_real_wrong_b.dat};
\addplot[barstyleVATRAM]
 table[x=set,y=vatram] {data/exp_comp_pe_real_wrong_b.dat};
\addplot[barstyleVATRAMnovar]
 table[x=set,y=vatramnovar] {data/exp_comp_pe_real_wrong_b.dat};
\addplot[barstyleMRSfast]
 table[x=set,y=mrsfast] {data/exp_comp_pe_real_wrong_b.dat};
\end{axis}
\end{tikzpicture}
&
\begin{tikzpicture}
\begin{axis}[
ybar,
ymin=0,
enlarge x limits=0.2,
bar width=0.25cm,
ylabel style={align=center},
yticklabel=
	\pgfmathparse{100*\tick}
	\pgfmathprintnumber{\pgfmathresult}\,\%,
yticklabel style={/pgf/number format/.cd,fixed,precision=2},
symbolic x coords={ERR967952},
xtick=data,
scaled ticks=false,
]
\addplot[barstyleBWA]
 table[x=set,y=bwa] {data/exp_comp_pe_real_wrong_c.dat};
\addplot[barstyleBWAedit]
 table[x=set,y=bwaedit] {data/exp_comp_pe_real_wrong_c.dat};
\addplot[barstyleBWAsw]
 table[x=set,y=bwasw] {data/exp_comp_pe_real_wrong_c.dat};
\addplot[barstyleBowtie]
 table[x=set,y=bowtie2] {data/exp_comp_pe_real_wrong_c.dat};
\addplot[barstyleVATRAM]
 table[x=set,y=vatram] {data/exp_comp_pe_real_wrong_c.dat};
\addplot[barstyleVATRAMnovar]
 table[x=set,y=vatramnovar] {data/exp_comp_pe_real_wrong_c.dat};
\addplot[barstyleMRSfast]
 table[x=set,y=mrsfast] {data/exp_comp_pe_real_wrong_c.dat};
\end{axis}
\end{tikzpicture}
\\
\begin{tikzpicture}
\begin{axis}[
ybar,
ymin=0.0,
enlarge x limits=0.2,
bar width=0.25cm,
ylabel style={align=center},
ylabel=runtime per \\ nucleotide (in \textmu s),
yticklabel style={/pgf/number format/.cd,fixed,precision=2},
symbolic x coords={ERR259389},
xtick=data,
scaled ticks=false,
]
\addplot[barstyleBWA]
 table[x=set,y=bwa] {data/exp_comp_pe_real_tbase_b.dat};
\addplot[barstyleBWAedit]
 table[x=set,y=bwaedit] {data/exp_comp_pe_real_tbase_b.dat};
\addplot[barstyleBWAsw]
 table[x=set,y=bwasw] {data/exp_comp_pe_real_tbase_b.dat};
\addplot[barstyleBowtie]
 table[x=set,y=bowtie2] {data/exp_comp_pe_real_tbase_b.dat};
\addplot[barstyleVATRAM]
 table[x=set,y=vatram] {data/exp_comp_pe_real_tbase_b.dat};
\addplot[barstyleVATRAMnovar]
 table[x=set,y=vatramnovar] {data/exp_comp_pe_real_tbase_b.dat};
\addplot[barstyleMRSfast]
 table[x=set,y=mrsfast] {data/exp_comp_pe_real_tbase_b.dat};
\end{axis}
\end{tikzpicture}
&
\begin{tikzpicture}
\begin{axis}[
ybar,
ymin=0,
enlarge x limits=0.2,
bar width=0.25cm,
ylabel style={align=center},
yticklabel style={/pgf/number format/.cd,fixed,precision=2},
symbolic x coords={ERR967952},
xtick=data,
scaled ticks=false,
]
\addplot[barstyleBWA]
 table[x=set,y=bwa] {data/exp_comp_pe_real_tbase_c.dat};
\addplot[barstyleBWAedit]
 table[x=set,y=bwaedit] {data/exp_comp_pe_real_tbase_c.dat};
\addplot[barstyleBWAsw]
 table[x=set,y=bwasw] {data/exp_comp_pe_real_tbase_c.dat};
\addplot[barstyleBowtie]
 table[x=set,y=bowtie2] {data/exp_comp_pe_real_tbase_c.dat};
\addplot[barstyleVATRAM]
 table[x=set,y=vatram] {data/exp_comp_pe_real_tbase_c.dat};
\addplot[barstyleVATRAMnovar]
 table[x=set,y=vatramnovar] {data/exp_comp_pe_real_tbase_c.dat};
\addplot[barstyleMRSfast]
 table[x=set,y=mrsfast] {data/exp_comp_pe_real_tbase_c.dat};
\end{axis}
\end{tikzpicture}
\end{tabular}
\ref*{leg:exp:rm:bars}
\end{center}
\caption{Real datasets with paired-end reads (see Table~\ref{tab:exp:real:overview}).
\emph{mrsFastUltra} crashes on both datasets.}
\label{fig:exp:real:pe}
\end{figure}
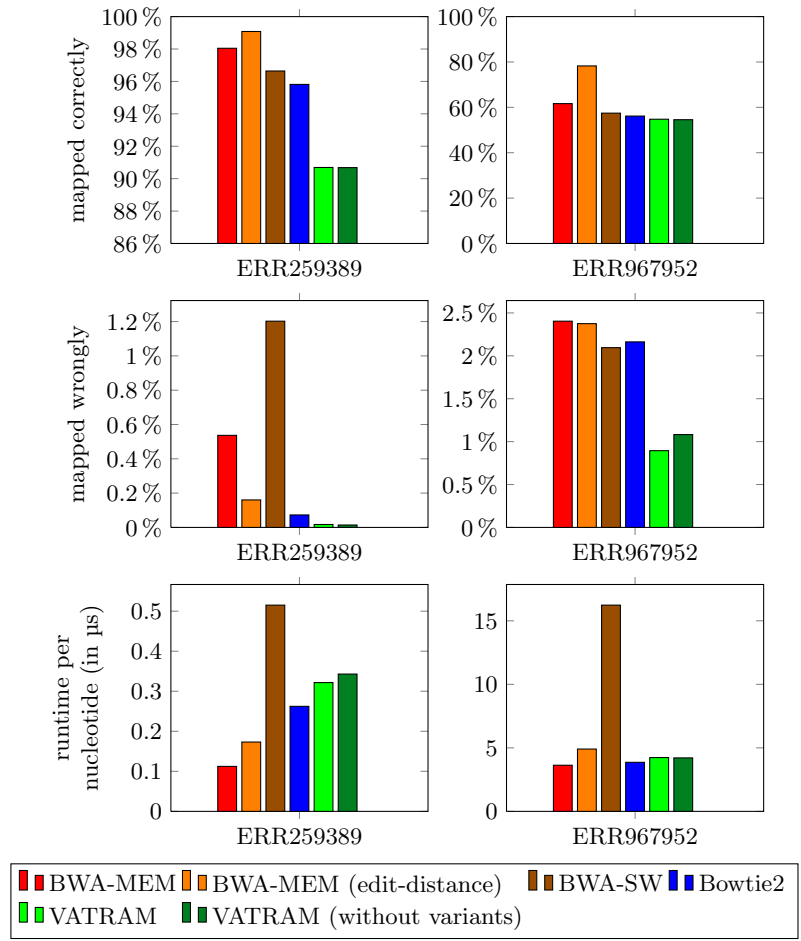

In Figure~\ref{fig:exp:real:pe} the results of the two paired-end data sets are shown. There are no values for \emph{mrsFastUltra}, because the program crashes for unknown reason. 

For the first dataset (\emph{ERR259389}) VATRAM achieves with 0.017\% an extremely low rate of wrongly mapped reads. The number of reads that Bowtie2 mapped wrongly are more than 4 times larger and the other read mappers produce even worse results according to this measure. On the other hand, only 90.7\% of the reads are mapped correctly by VATRAM. The other read mappers achieve 95.8\% or more; so many reads remain unmapped when using VATRAM.

In the dataset \emph{ERR967952} the ratios of correctly mapped reads are almost equal for VATRAM, Bowtie2, BWA-SW and BWA-MEM in standard configuration. However, BWA-MEM in edit-distance configuration is able to map about 17\%-points more reads correctly than the other read mappers.
On the other hand the number of wrongly mapped reads of BWA-MEM (in both configuraitons), Bowtie2 and BWA-SW is always twice as large as the number of reads that VATRAM mapped wrongly.
Which read mapper is better depends on the application: If you want to have many correctly mapped reads, it makes sense to use BWA-MEM in edit-distance configuration, if you want to minimize the number of wrongly mapped reads, it is better to use VATRAM.
The runtime of VATRAM in this dataset is comparable with the runtime of Bowtie2 and BWA-MEM. This is interesting, because the alignment process of VATRAM is usually much more time consuming due to the consideration of variants.

For both data sets there is only a slight benefit when the variants are added to the index.
According to the number of correctly mapped reads, there is no change visible in Figure~\ref{fig:exp:real:pe}.
The ratio of wrongly mapped reads decreases for the data set \emph{ERR967952} by about 17\%.
The runtime was reduced by approximately 6\% in the dataset \emph{ERR259389}.
For the other data sets there was no significant change.

\subsubsection{Real reads with variable length}
\label{sec:exp:real:var}

\begin{figure}[tp]
\pgfplotsset{footnotesize,height=4.6cm}
\begin{center}
\setlength{\tabcolsep}{0mm}
\begin{tabular}{rr}
\pgfplotsset{width=6.5cm}
\begin{tikzpicture}
\begin{axis}[
ybar,
ymin=0.86,ymax=1,
enlarge x limits=0.5,
bar width=0.25cm,
ylabel style={align=center},
ylabel=mapped correctly,
yticklabel=
	\pgfmathparse{100*\tick}
	\pgfmathprintnumber{\pgfmathresult}\,\%,
yticklabel style={/pgf/number format/.cd,fixed,precision=2},
symbolic x coords={DRR003760, SRR003174},
xtick=data,
scaled ticks=false,
]
\addplot[barstyleBWA]
 table[x=set,y=bwa] {data/exp_comp_long_real_corr_abc.dat};
\addplot[barstyleBWAedit]
 table[x=set,y=bwaedit] {data/exp_comp_long_real_corr_abc.dat};
\addplot[barstyleBWAsw]
 table[x=set,y=bwasw] {data/exp_comp_long_real_corr_abc.dat};
\addplot[barstyleBowtie]
 table[x=set,y=bowtie2] {data/exp_comp_long_real_corr_abc.dat};
\addplot[barstyleVATRAM]
 table[x=set,y=vatram] {data/exp_comp_long_real_corr_abc.dat};
\addplot[barstyleVATRAMnovar]
 table[x=set,y=vatramnovar] {data/exp_comp_long_real_corr_abc.dat};
\end{axis}
\end{tikzpicture}
&
\pgfplotsset{width=4.5cm}
\begin{tikzpicture}
\begin{axis}[
ybar,
ymin=0.0,ymax=1,
enlarge x limits=0.25,
bar width=0.25cm,
ylabel style={align=center},
yticklabel=
	\pgfmathparse{100*\tick}
	\pgfmathprintnumber{\pgfmathresult}\,\%,
yticklabel style={/pgf/number format/.cd,fixed,precision=2},
symbolic x coords={SRX533609},
xtick=data,
scaled ticks=false,
]
\addplot[barstyleBWA]
 table[x=set,y=bwa] {data/exp_comp_long_real_corr_d.dat};
\addplot[barstyleBWAedit]
 table[x=set,y=bwaedit] {data/exp_comp_long_real_corr_d.dat};
\addplot[barstyleBWAsw]
 table[x=set,y=bwasw] {data/exp_comp_long_real_corr_d.dat};
\addplot[barstyleBowtie]
 table[x=set,y=bowtie2] {data/exp_comp_long_real_corr_d.dat};
\addplot[barstyleVATRAM]
 table[x=set,y=vatram] {data/exp_comp_long_real_corr_d.dat};
\addplot[barstyleVATRAMnovar]
 table[x=set,y=vatramnovar] {data/exp_comp_long_real_corr_d.dat};
\end{axis}
\end{tikzpicture}
\\
\pgfplotsset{width=6.5cm}
\begin{tikzpicture}
\begin{axis}[
ybar,
ymin=0.0,
enlarge x limits=0.5,
bar width=0.25cm,
ylabel style={align=center},
ylabel=mapped wrongly,
yticklabel=
	\pgfmathparse{100*\tick}
	\pgfmathprintnumber{\pgfmathresult}\,\%,
yticklabel style={/pgf/number format/.cd,fixed,precision=2},
symbolic x coords={DRR003760, SRR003174},
xtick=data,
scaled ticks=false,
]
\addplot[barstyleBWA]
 table[x=set,y=bwa] {data/exp_comp_long_real_wrong_abc.dat};
\addplot[barstyleBWAedit]
 table[x=set,y=bwaedit] {data/exp_comp_long_real_wrong_abc.dat};
\addplot[barstyleBWAsw]
 table[x=set,y=bwasw] {data/exp_comp_long_real_wrong_abc.dat};
\addplot[barstyleBowtie]
 table[x=set,y=bowtie2] {data/exp_comp_long_real_wrong_abc.dat};
\addplot[barstyleVATRAM]
 table[x=set,y=vatram] {data/exp_comp_long_real_wrong_abc.dat};
\addplot[barstyleVATRAMnovar]
 table[x=set,y=vatramnovar] {data/exp_comp_long_real_wrong_abc.dat};
\end{axis}
\end{tikzpicture}
&
\pgfplotsset{width=4.5cm}
\begin{tikzpicture}
\begin{axis}[
ybar,
ymin=0.0,
enlarge x limits=0.25,
bar width=0.25cm,
ylabel style={align=center},
yticklabel=
	\pgfmathparse{100*\tick}
	\pgfmathprintnumber{\pgfmathresult}\,\%,
yticklabel style={/pgf/number format/.cd,fixed,precision=2},
symbolic x coords={SRX533609},
xtick=data,
scaled ticks=false,
]
\addplot[barstyleBWA]
 table[x=set,y=bwa] {data/exp_comp_long_real_wrong_d.dat};
\addplot[barstyleBWAedit]
 table[x=set,y=bwaedit] {data/exp_comp_long_real_wrong_d.dat};
\addplot[barstyleBWAsw]
 table[x=set,y=bwasw] {data/exp_comp_long_real_wrong_d.dat};
\addplot[barstyleBowtie]
 table[x=set,y=bowtie2] {data/exp_comp_long_real_wrong_d.dat};
\addplot[barstyleVATRAM]
 table[x=set,y=vatram] {data/exp_comp_long_real_wrong_d.dat};
\addplot[barstyleVATRAMnovar]
 table[x=set,y=vatramnovar] {data/exp_comp_long_real_wrong_d.dat};
\end{axis}
\end{tikzpicture}
\\
\pgfplotsset{width=6.5cm}
\begin{tikzpicture}
\begin{axis}[
ybar,
ymin=0.0,
enlarge x limits=0.5,
bar width=0.25cm,
ylabel style={align=center},
ylabel=runtime per\\ nucleotide (in \textmu s),
yticklabel style={/pgf/number format/.cd,fixed,precision=2},
symbolic x coords={DRR003760, SRR003174},
xtick=data,
scaled ticks=false,
]
\addplot[barstyleBWA]
 table[x=set,y=bwa] {data/exp_comp_long_real_tbase_abc.dat};
\addplot[barstyleBWAedit]
 table[x=set,y=bwaedit] {data/exp_comp_long_real_tbase_abc.dat};
\addplot[barstyleBWAsw]
 table[x=set,y=bwasw] {data/exp_comp_long_real_tbase_abc.dat};
\addplot[barstyleBowtie]
 table[x=set,y=bowtie2] {data/exp_comp_long_real_tbase_abc.dat};
\addplot[barstyleVATRAM]
 table[x=set,y=vatram] {data/exp_comp_long_real_tbase_abc.dat};
\addplot[barstyleVATRAMnovar]
 table[x=set,y=vatramnovar] {data/exp_comp_long_real_tbase_abc.dat};
\end{axis}
\end{tikzpicture}
&
\pgfplotsset{width=4.5cm}
\begin{tikzpicture}
\begin{axis}[
ybar,
ymin=0.0,
enlarge x limits=0.25,
bar width=0.25cm,
ylabel style={align=center},
yticklabel style={/pgf/number format/.cd,fixed,precision=2},
symbolic x coords={SRX533609},
xtick=data,
scaled ticks=false,
]
\addplot[barstyleBWA]
 table[x=set,y=bwa] {data/exp_comp_long_real_tbase_d.dat};
\addplot[barstyleBWAedit]
 table[x=set,y=bwaedit] {data/exp_comp_long_real_tbase_d.dat};
\addplot[barstyleBWAsw]
 table[x=set,y=bwasw] {data/exp_comp_long_real_tbase_d.dat};
\addplot[barstyleBowtie]
 table[x=set,y=bowtie2] {data/exp_comp_long_real_tbase_d.dat};
\addplot[barstyleVATRAM]
 table[x=set,y=vatram] {data/exp_comp_long_real_tbase_d.dat};
\addplot[barstyleVATRAMnovar]
 table[x=set,y=vatramnovar] {data/exp_comp_long_real_tbase_d.dat};
\end{axis}
\end{tikzpicture}
\end{tabular}
\ref*{leg:exp:rm:bars}
\end{center}
\caption{Real reads with variable length.
Information about the data sets can be found in Table~\ref{tab:exp:real:overview}.
There are no results for \emph{mrsFast-Ultra}, because it is not able to map reads with variable length.}
\label{fig:exp:real:var}
\end{figure}
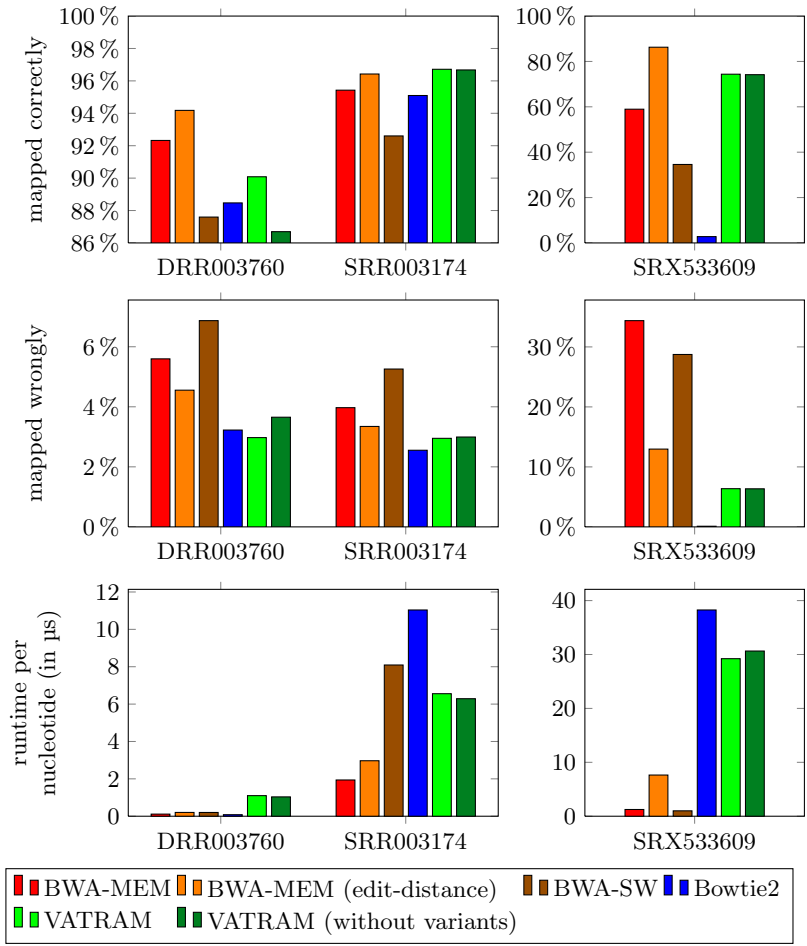

So far we only analyzed data sets where all reads have the same length.
However, sequencing machines like ``454 GS'' FLX or PacBio produce reads with varying length. In Figure~\ref{fig:exp:real:var} the results for three such data sets are shown.

The data set \emph{DRR003760} is interesting, because considering the variants leads to clearly better results for VATRAM.
The ratio of correctly mapped reads increases from 86.7\% to 90.1\%.
Such a high improvement of 3.4\%-points was not measured in other datasets, neither for the synthetic nor for the real data sets.
The reason for this is that the reads of the \emph{DRR003760} are located in the human HLA region which contains much variants in a very small area.
Therefore, there is a great benefit, if the known variants are added to VATRAMs index.
However, the ratio of correctly mapped reads is still lower than the corresponding ratio of BWA-MEM (92.3\% in standard respectively 94.2\% in edit-distance configuration).

The data set \emph{SRR003174} shows that VATRAM is able to produce better results than BWA-MEM: Not only is VATRAMs ratio of wrongly mapped reads smaller than the ratio of BWA-MEM (in both configurations), but VATRAM also maps more reads to the correct position, even for the edit distance configuration of BWA-MEM which always leads to the highest number of correctly mapped read in the other experiments.

The data sets \emph{DRR003760} and \emph{SRR003174} contain reads with an moderate read length of 180 respectively 565 nucleotides on average. The last column in figure~\ref{fig:exp:real:var} shows the results for a read data set generated by a PacBio sequencing machine whose reads are much longer (\emph{SRX533609}). VATRAMs aligner was not constructed to align such long reads with $10\,000$ nucleotides or even more. Therefore its running time is more than 20 times longer  than the running of BWA-MEM in standard configuration. Only Bowtie2 needs even more time than VATRAM, since it was not designed for such long reads either. According to the number of correctly and wrongly mapped reads VATRAM performs clearly better than BWA-MEM in standard configuration. However, if the edit distance configuration of BWA-MEM is used (which makes more sense, since the edit distance is used as measure to determine whether a read is mapped correctly), then the ratio of correctly  mapped reads is 12\%-points better than VATRAMs ratio. However, also the ratio of wrongly mapped reads is 6.6\%-points greater than the corresponding ratio of VATRAM. Which read mapper is better in this case depends on the application. VATRAM achieves the best precision, however the recall of BWA-MEM in edit-distance configuration is better than VATRAMs recall.

\section{Discussion and conclusion}
\label{sec:discussion}

In the last decades many read mappers were developed, so there is the question why one should develop another read mapper.
Most of the commonly used read mappers do not use knowledge about known variants.
However, there are many differences between the reference genome and the genome of one individual.
By now about 150 millions of variants are known\footnote{\url{https://www.ncbi.nlm.nih.gov/dbvar/content/org\_summary}, access date: 20th January 2017.} whose usage may improve the mapping process significantly.
Common read mappers (like BWA or Bowtie2) treat these variants similarly to sequencing errors and thus may have problems to map reads containing known variants to the correct position.
Therefore we developed a new read mapper called VATRAM based on min-hasing that is able to consider known variants.

In section~\ref{sec:exp} we compared VATRAM with other read mappers.
In most cases BWA-MEM in edit-distance configuration performs best according to the ratio of correctly mapped reads.
For the data set \emph{SRR003174}, VATRAM was the best read mapper.
An additional strength of VATRAM is the very low ratio of wrongly mapped reads. It was usually much better than the ratio of Bowtie2 or BWA.

We tested not only common, non-variant tolerant read mappers like BWA or Bowtie2, but also the variant tolerant read mapper \emph{mrsFastUltra}.
However, its results were not convincing according to the very low rate of correctly mapped reads.
Furthermore, it failed to handle all five tested real data sets.

VATRAM needs more time than BWA or Bowtie2 to process a data set.
This is not surprising, since calculating a variant tolerant alignment is more complex and therefore more time consuming than a usual alignment that is for example performed by Bowtie2 and BWA.

We executed VATRAM with and without the information about the known variants.
The benefit of considering the variants is usually small, because only a small ratio of the reads contain variants.
However, for the data set \emph{DRR003760} which contains reads from a region with many variants in a small area the mapping quality of VATRAM improved significantly after adding the known variants to the index.

All in all, VATRAM is a competitive read mapper in comparison to BWA and Bowtie2.
Currently VATRAM is implemented as a prototype, and certainly further optimizations in terms of speed and memory usage are possible.
An open question is in how many individuals a variant must appear such that adding this variant to the index leads to better results.
This question could also be analyzed theoretically (using lemma~\ref{lemma:minhash}), independent from the details of VATRAMs implementation.

\bibliographystyle{splncs03}
\bibliography{literature}

\begin{thebibliography}{10}
\providecommand{\url}[1]{\texttt{#1}}
\providecommand{\urlprefix}{URL }

\bibitem{Berlin2015}
Berlin, K., Koren, S., Chin, C.S., Drake, J.P., Landolin, J.M., Phillippy,
  A.M.: Assembling large genomes with single-molecule sequencing and
  locality-sensitive hashing. Nature biotechnology  33(6),  623--630 (2015)

\bibitem{Broder1997}
Broder, A.Z.: On the resemblance and containment of documents. In: Compression
  and Complexity of Sequences (SEQUENCES'97). pp. 21--29. IEEE (1997)

\bibitem{Broder1998}
Broder, A.Z., Charikar, M., Frieze, A.M., Mitzenmacher, M.: Min-wise
  independent permutations. In: Proceedings of the 30th annual {ACM} symposium
  on Theory of computing (STOC). pp. 327--336. ACM (1998)

\bibitem{Buhler2001}
Buhler, J.: Efficient large-scale sequence comparison by locality-sensitive
  hashing. Bioinformatics  17(5),  419--428 (2001)

\bibitem{Xor2013}
Casperson, M.: Minhash for dummies.
  \url{http://matthewcasperson.blogspot.de/2013/11/minhash-for-dummies.html}
  (November 2013)

\bibitem{PacBio2009}
Eid, J., Fehr, A., Gray, J., Luong, K., Lyle, J., Otto, G., Peluso, P., Rank,
  D., Baybayan, P., Bettman, B., Bibillo, A., Bjornson, K., Chaudhuri, B.,
  Christians, F., Cicero, R., Clark, S., Dalal, R., deWinter, A., Dixon, J.,
  Foquet, M., Gaertner, A., Hardenbol, P., Heiner, C., Hester, K., Holden, D.,
  Kearns, G., Kong, X., Kuse, R., Lacroix, Y., Lin, S., Lundquist, P., Ma, C.,
  Marks, P., Maxham, M., Murphy, D., Park, I., Pham, T., Phillips, M., Roy, J.,
  Sebra, R., Shen, G., Sorenson, J., Tomaney, A., Travers, K., Trulson, M.,
  Vieceli, J., Wegener, J., Wu, D., Yang, A., Zaccarin, D., Zhao, P., Zhong,
  F., Korlach, J., Turner, S.: Real-time {DNA} sequencing from single
  polymerase molecules. Science  323(5910),  133--138 (2009)

\bibitem{fmindex}
Ferragina, P., Manzini, G.: Indexing compressed text. Journal of the {ACM}
  52(4),  552--–581 (2005)

\bibitem{mrsFastUltra2014}
Hach, F., Sarrafi, I., Hormozdiari, F., Alkan, C., Eichler, E.E., Sahinalp,
  S.C.: {mrsFAST-Ultra}: a compact, {SNP}-aware mapper for high performance
  sequencing applications. Nucleic Acids Research  42(Webserver-Issue),
  494--500 (2014)

\bibitem{PG583}
Kramer, B., Quedenfeld, J., Schrinner, S., Bargull, M., Benadjemia, K.,
  Stricker, J., Losch, D.: {V}{A}{T}{R}{A}{M} -- {V}{A}riant {T}olerant
  {R}e{A}d {M}apper. Tech. rep., Project Group PG583, Computer Science, TU
  Dortmund, Germany (2015)

\bibitem{Bowtie2012}
Langmead, B., Salzberg, S.L.: Fast gapped-read alignment with {Bowtie 2}.
  Nature methods  9(4),  357--359 (2012)

\bibitem{BWA2013}
Li, H.: Aligning sequence reads, clone sequences and assembly contigs with
  {BWA-MEM}. arXiv preprint arXiv:1303.3997  (2013)

\bibitem{BWA2009}
Li, H., Durbin, R.: Fast and accurate short read alignment with
  {Burrows--Wheeler transform}. Bioinformatics  25(14),  1754--1760 (2009)

\bibitem{BWA2010}
Li, H., Durbin, R.: Fast and accurate long-read alignment with {Burrows-Wheeler
  transform}. Bioinformatics  26(5),  589--595 (2010)

\bibitem{Metzker2010}
Metzker, M.L.: Sequencing technologies---the next generation. Nature Reviews
  Genetics  11(1),  31--46 (2010)

\bibitem{Navarro2016}
Navarro, G.: Compact Data Structures: A Practical Approach. Cambridge
  University Press, Cambridge (009 2016)

\bibitem{MA}
Quedenfeld, J.: Variantentolerantes Readmapping durch Locality Sensitive
  Hashing. Master's thesis, Computer Science XI, TU Dortmund, Germany (2016)

\bibitem{Ukkonen1985}
Ukkonen, E.: Finding approximate patterns in strings. Journal of Algorithms
  6(1),  132 -- 137 (1985)

\end{thebibliography}
\clearpage




\end{document}